\documentclass[preprint]{aastex631}
\usepackage{fancyhdr}
\usepackage{ulem}
\received{??, 2023}
\revised{??, 2024}
\accepted{\today}
\submitjournal{ApJ}

\shorttitle{Statistical Relationships between HD and disk gas mass}
\shortauthors{Seo et al.}

\begin{document}

\title{Retrievals of Protoplanetary Disk Parameters using Thermochemical Models: \\ I. Disk Gas Mass from Hydrogen Deuteride Spectroscopy}

\correspondingauthor{Young Min Seo}
\email{youngmin.seo@jpl.nasa.gov}

\author[0000-0003-2122-2617]{Young Min Seo}
\affiliation{Jet Propulsion Laboratory, California Institute of Technology, 4800 Oak Grove Dr. Pasadena, CA, 91109, USA}

\author[0000-0001-6124-5974]{Karen Willacy}
\affiliation{Jet Propulsion Laboratory, California Institute of Technology, 4800 Oak Grove Dr. Pasadena, CA, 91109, USA}

\author[0000-0001-5966-837X]{Geoffrey Bryden}
\affiliation{Jet Propulsion Laboratory, California Institute of Technology, 4800 Oak Grove Dr. Pasadena, CA, 91109, USA}

\author[0000-0002-0500-4700]{Dariusz C. Lis}
\affiliation{Jet Propulsion Laboratory, California Institute of Technology, 4800 Oak Grove Dr. Pasadena, CA, 91109, USA}

\author[0000-0002-6622-8396]{Paul F. Goldsmith}
\affiliation{Jet Propulsion Laboratory, California Institute of Technology, 4800 Oak Grove Dr. Pasadena, CA, 91109, USA}

\author[0000-0001-7552-1562]{Klaus M. Pontoppidan}
\affiliation{Jet Propulsion Laboratory, California Institute of Technology, 4800 Oak Grove Dr. Pasadena, CA, 91109, USA}

\author[0000-0002-1363-2301]{Wing-Fai Thi}
\affiliation{Max Planck Institute for Extraterrestrial Physics, Giessenbachstrasse 1, D-85748 Garching, Germany}

\begin{abstract}

We discuss statistical relationships between the mass of protoplanetary disks, the hydrogen deuteride (HD) line { flux}, and the dust spectral energy distribution (SED) determined using 3000 ProDiMo disk models. The models have 15 free parameters describing { the} disk physical properties, the central star, and the local radiation field. The sampling of physical parameters is done using a Monte Carlo approach to evaluate the probability density functions of observables as a function of physical parameters. We find that the { mass-averaged} HD fractional abundance is almost constant even though the UV flux varies by several orders of magnitude. Probing the statistical relation between the physical quantities and the HD flux, we find that low-mass (optically thin) disks display a tight correlation between the average disk gas temperature and HD line flux, while massive disks show no such correlation. We demonstrate that the central star luminosity, disk size, dust size distribution, and HD flux may be used to determine the disk gas mass within a factor of three. We also find that the far-IR and sub-mm/mm SEDs and the HD flux may serve as strong constraints for determining the disk gas mass to within a factor of two. If the HD lines are fully spectrally resolved ($R\gtrsim 1.5\times10^6, \Delta v=0.2~\rm km\,s^{-1}$), the 56 $\mu$m and 112 $\mu$m HD line profiles alone may constrain the disk gas mass to within a factor of two.

\end{abstract}


\section{Introduction} \label{sec:intro}

Planet-forming disks are a critical step in the formation of planetary systems. They set the initial conditions of planet formation \citep[e.g.,][]{cleeves14,pontoppidan14,podio19,tobin23}, and their detailed processes are thought to determine the characteristics and demographics of planetary systems. Disks are found around all types of young stars except O-type stars \citep[e.g.,][]{odell94,evens09,williams11,beltran14,cesaroni17}, suggesting that planet formation may occur throughout the Universe. 

{ Multi-wavelength s}patially resolved images obtained using large telescopes such as ALMA, Gemini, Keck, and VLT have revolutionized our view of planet-forming disks \citep[e.g.,][]{defrere12,andrews18,rich22}. While all disks have flattened circular geometry, the high spatial resolution provided by these large telescopes has revealed the complex nature and exceptional diversity of disk structures. ALMA images, in particular, show gaps, spiral arms, asymmetries, and evidence of dust settling in many disks. These observational results demonstrate that { disk evolution involves} multiple physical and chemical processes and are complicated factories of diverse planets. However, the challenge lies in following the processes given the very diverse characteristics of planet-forming disks.

Currently, two frameworks exist for understanding planet formation depending on whether the disk is gravitationally stable or not. For a stable disk, planets are thought to form by core-accretion \citep[e.g.,][]{pollack96,rice03,ida04} where small dust grains grow into protoplanets by first forming pebbles, then planetesimals, and finally protoplanets through collisions.  Protoplanets larger than a certain mass may undergo runaway gas accretion and evolve into gas giants. The details of the accretion steps are still a matter of debate, with several mechanisms having been suggested for different time epochs and disk physical conditions \citep[e.g.,][]{johansen14}. In the case of an unstable disk, planets can potentially form by direct gravitational collapse \citep[e.g.,][]{boss97,rice22}, allowing gas giant planets to form relatively quickly, whereas small planets still form according to the core-accretion model. In both cases, there are many details that are not yet well understood or constrained by observations, even in the case of our own solar system, and we do not yet fully understand why there is such a wide diversity of exoplanetary systems. 

To constrain the key processes in planet formation, we must start by finding the observables that are sensitive to the parameters or processes of interest. This can be challenging, as a good tracer needs to be tightly correlated to a specific physical process/parameter in order to minimize the uncertainty of the retrieved physical parameter \citep[e.g.,][]{maret99,mcnabb13,pineda19}. Given the physical and chemical complexity of protoplanetary disks, it can be difficult to identify tracers that correlate in simple ways with fundamental parameters and processes.

This work describes a statistical approach to understanding the relationship between observables and physical parameters in protoplanetary disks. Using 3000 protoplanetary disk models including chemistry and radiative transfer, we have probed the statistical correlations between the observables and physical parameters (science retrievals). The work also analyzes the limitations and uncertainties of using specific observables to constrain disks' physical parameters. In this first paper of the series, we focus on the correlation between hydrogen deuteride (HD) lines and the disk gas mass.   

To understand the first steps in planet formation, we need to know the initial dust and gas distributions in a disk. Of the two, observing the dust continuum emission is significantly less demanding in terms of telescope sensitivity since the continuum { bandwidths} are much wider than the gas emission lines. Dust continuum observations from ALMA and the VLA, along with scattered light images from large optical and near-IR observatories, have provided a major data stream for studying planet formation \citep[e.g.,][]{andrews18,tobin19,rich22}. In contrast, we are missing critical information regarding the gas component, which has fairly limited observations to date \citep[e.g.,][]{oberg21,teague22}. Initially, the gas comprises 99\% of the disk mass, but it is expected to dissipate over time. The total mass and spatial distribution of the gas, the dissipation timescale, outflows, and gas turbulence are thought to be key for constraining the planet formation process. Unfortunately, quantifying such information has been challenging due to the lack of an optimal tracer for the gas component.

Molecular hydrogen (H$_2$) is the dominant component of the gas in disks. However, observing bulk H$_2$ directly in disks is challenging since it is a homonuclear molecule without a permanent electric dipole moment and its rotational levels are connected only by weak quadrupole transitions. Although commonly observed in the ultraviolet (UV) from the uppermost layers of disks \citep{France12}, the weakness of the rotational H$_2$ transitions, coupled with their high energies, mean that it is essentially not emissive from the vast majority of the disk gas mass \citep{Bitner08,Carmona08,fedele13}. To probe this cold gas component emission from carbon monoxide (CO) isotopologues is traditionally used \citep{Dutrey96,Ansdell16}, but the conversion from CO to total disk mass is dependent on complex { physical and chemical} processes leading to large uncertainties \citep{Bruderer12, Kama16, Ruaud22, Pascucci23}. A more recent alternate tracer of disk gas mass is singly deuterated molecular hydrogen \citep[(HD),][]{Bergin13,McClure16}.    

\citet{trapman17} and \citet{Calahan21} demonstrated the potential of HD as a robust tracer of the total gas mass in protoplanetary disks. They modeled the chemistry (including { deuteration}) in disks and simulated the HD line emission. They showed that the HD flux has a monotonic correlation with the total disk gas mass for typical protoplanetary disks. \citet{kama20} also studied the correlation between the HD flux and the total gas mass in disks around more massive stars ($>2\,M_{\odot}$). They found that the overall HD abundance across disks is almost constant, similar to disks around low-mass stars, and the HD flux also shows a monotonic correlation with the total gas mass. While these studies provide a basic understanding of the relation between HD line flux and disk mass, they are limited to a relatively narrow range of model parameters and assume that many disk properties (e.g., inclinations, disk size, central star luminosity, UV flux, etc.) are known from other observations (e.g., ALMA). The statistical relationship between the HD flux and the disk gas structure when disk properties are not well-known has not been studied { previously}. Thus, the limitations of using the HD emission for probing disk gas { distribution} and the resulting uncertainty of the mass retrievals from the HD emission are still to be fully quantified.

This work investigates the statistical relationship between 56 and 112 $\mu$m HD emission and the disk gas mass over a much larger parameter space than that explored in \citet{trapman17} and \citet{kama20}. We analyze the uncertainty of the disk gas mass estimates using the HD flux { as a function of} various physical configurations of the disks and the stellar properties. We aim to establish what essential minimal information must be known for HD to { accurately} determine the gas mass in disks. We find an appreciable scatter between the HD flux and the disk gas mass if no other information about the disk is known. { However, we find that the HD emission may be a solid measure of the disk gas mass even when only three observables (the spectral energy distribution (SED) and the disk inclination along with HD flux) are obtained, which is significantly less number of the known observables/information considered in \citet{trapman17} and \citet{Calahan21}.} 

We describe the details of the disk models in Section \S2. The statistical relationship between the HD flux and the disk gas mass and further analysis are presented in Section \S3. Physical parameters that affect the HD fluxes are also explained in Section \S3, along with the minimal essential information required to determine the disk gas mass. In Section \S4, we discuss the limitations of the models, and we summarize the findings in Section \S5.    

\section{Modeling planet-forming disks} \label{sec:mod}

We use the Protoplanetary Disk Model (ProDiMo) code \citep{woitke09,kamp10,thi20} to model the chemistry and observables in disks. The ProDiMo code includes multiple physics and chemistry modules, including radiative transfer, hydrostatic equation solver, and two-phase chemistry (gas and grain). The disk structure is defined either parametrically or by assuming hydrostatic equilibrium, and the temperature distribution is then determined by balancing the heating and cooling of the gas and dust.  The code contains multiple options for chemistry networks (e.g., UMIST, KIDA, OSU, and pre-compiled grain chemistry) and may simulate either equilibrium or time-dependent chemistry according to the disk's physical structure. Finally, the code delivers observables using a radiative transfer module from the disks' calculated physical and chemical structures. All of the steps are done self-consistently based on the given parameters, and the code delivers a unique set of observables from a given set of physical and chemical parameters.    

We focus on parametric disks, although they may not always represent self-consistent, physical solutions. There are two reasons for this: (1) disks are highly dynamic structures with dust-gas interactions and may not actually be in thermal hydrostatic equilibrium \citep[e.g.,][]{lesur22}, and (2) disk turbulence, which together with thermal pressure, supports the disk vertical structure against self-gravity is not very well known, and may be generated by multiple mechanisms  \citep[e.g.,][]{lyra19,gole20} that cannot be simulated in the ProDiMo code. A hydrostatic disk that is purely supported by thermal pressure may not be realistic. Instead, parametric models { may describe disk structure directly} as free parameters { (see Appendix)}, including the flaring index, disk vertical scale height, and power law index of the radial disk surface density profile, { which may reflect non-equilibrium processes}.

The density of a parametric disk is defined by the parameters listed in Table 1 in \citet{woitke10}. Additional model parameters include the disk inner and outer radii, dust/gas mass ratio, disk vertical viscosity, and an option for dust settling description. We assume static parametric disks except for dust settling. Any motions such as radial and vertical transport are not considered in this study. While the dynamics may affect the chemistry and observables \citep[e.g., from pebble drift;][]{Pinilla12, Kalyaan21}, we focus here on static disks because it is not computationally feasible to generate a sufficiently large database of models that include a full description of coupled dynamics and chemistry in a reasonable time frame. 

Dust properties can also be configured in the code. The dust grains play an important role in the radiative transfer calculations, impacting disk temperature, opacity, and chemistry. In this study, we use dust grains with 30\% ice, 40\% astro-silicate, and 30\% carbonaceous grains with 80\% porosity, a composition intermediate between bare refractory grains \citep[e.g.,][]{draine03} and cold icy grains \citep[e.g.,][]{potapov20}. This composition is only used { to} estimate opacity and calculate the temperature structure of the disk. The grain size distribution in the code is described by three parameters (the minimum and maximum dust sizes and the power index of dust size distribution), which are varied as free parameters in this study. { The dust size distribution is sampled using 100 bins.} The opacity of dust grains as a function of wavelength is estimated using Mie theory (Miex, \citealt{wolf04}). We use the dust settling description from \citet{dubrulle95}. This is an exact physical description of dust settling for a given vertical viscosity. However, the vertical viscosity of the disk is not well known and depends on the specific mechanism of vertical turbulence in disks. Thus, we chose to keep the disk viscosity as a free parameter in this study.  

The radiative transfer is carried out using a ray-tracing radiative transfer module, which is part of ProDiMo. We consider two sources of radiation: the central star and the interstellar radiation field (ISRF). The central star luminosity is determined by following a stellar evolutionary model { \citep{somers20}} for a given stellar mass and age. The stellar mass and the age are chosen randomly within a range (0.1 Myr $-$ 50 Myr). { The UV radiation from the central star is considered as a fraction of the stellar bolometric luminosity with a spectral power index fixed to 1.3.} The interstellar radiation field is implemented as vertically impinging radiation onto the disk and is in the unit of G$_\circ$ (1.3 Habing fields, \citet{habing68,mathis83}). X-rays are not considered in this study since version 2.0 of the ProDiMo code used in this study did not fully implement a coupling between X-rays and deuterium chemistry. However, we compared our results to the models with X-ray radiative transfer but without { deuteration}. We found that X-rays have only a minor impact on a statistical study but may play an important role in finding the best fit for a specific target. We describe the self-shielding of H$_2$, HD, and H using the treatment of \citet{wolcott-green11}. Heating due to gas accretion is not considered in this study to keep the number of free parameters within a practical range so that we can carry out a large number of models on a reasonable time scale.

We use a time-dependent two-phase chemistry (gas and dust surface) network included in the ProDiMo code. We have 14 elements in the chemical network, including deuterium \citep{thi20}. The atomic elements and the initial atomic abundance are shown in Table \ref{tab:el}. The chemical network involves 304 atomic and molecular species with 215 gas species and 89 ice species. The chemical reaction network includes 4450 reactions involving gas species, ice species, ions, and photons. All the chemistry is simulated using the gas and dust grain temperature profiles estimated from the radiative transfer calculations. The UV field is also calculated during radiative transfer calculations.

The initial disk composition is assumed to be inherited from a molecular cloud core. We do not consider the possibility that the abundances are completely ``reset'' (i.e., molecules created in the parent molecular cloud core are destroyed and the chemistry in the disk starts from elemental abundances) during the disk formation. While the degree of inheritance is still under debate \citep{pontoppidan14}, there are signatures of inheritance from a dense molecular clouds core to a disk in the nitrogen fractionation and NH$_3$ spin temperatures in comets \citep{mumma11} and the HDO/H$_2$O ratio of V883 Ori, which is undergoing an accretion outburst \citep{tobin23}. Thus, we assume that the gas and dust grains are inherited from a molecular cloud core. The ProDiMo code has a consistent chemistry module simulating the chemistry of a molecular cloud core, which is a single-point chemistry for a given density and temperature. We simulated chemistry at 10$^6$ cm$^{-3}$ at 10 K and adopted the chemical abundance at 1 Myr as the initial chemical abundance of disks.
\begin{deluxetable*}{cc}
\tablecaption{Elemental Abundance\label{tab:el}}
\tablewidth{0pt}
\tablehead{
\colhead{Element} & \colhead{Abundance$^a$} 
}
\startdata
H   & 12.000  \\
He  & 10.954  \\
D   & 7.1883  \\
C   & 8.1293  \\
N   & 7.9610  \\
O   & 8.5040  \\
Ne  & 7.8386  \\
Na  & 2.2998  \\
Mg  & 3.1759  \\
Si  & 3.9954  \\
S   & 4.9048  \\
Ar  & 6.1759  \\
Fe  & 3.4607  \\
PAH & 3.4485  
\enddata
\tablecomments{$^a$The abundance values are shown as Log$_{10}$(X/H)+12.}
\end{deluxetable*}
\begin{deluxetable*}{ccc}
\tablecaption{Model parameters\label{tab:para}}
\tablewidth{0pt}
\tablehead{
\colhead{Parameter} & \colhead{Range} & \colhead{Unit} 
}
\startdata
Stellar mass, M$_*$                  & 0.23 $-$ 2.2       & M$_\odot$                           \\
Central star luminosity, L$_*$       & 0.014 $-$ 43       & L$_\odot$                            \\
UV fraction, L$_{\rm UV}$/L$_*$      & 10$^{-6}$ $-$ 0.1  & -                           \\
Cosmic-ray flux, F$_{\rm CR}$        & 10$^{-20}$ $-$ 2 $\times$ 10$^{-16}$ & cm$^{-2}$  \\
ISRF, F$_{\rm ISM}$                  & 0.01 $-$ 10       & G$_0$                         \\
dust/gas mass density ratio, $\rho_d$/$\rho_g$    & 0.00625 $-$ 0.25   & -                                 \\
Ratio of disk mass to stellar mass, M$_{\rm disk}$/M$_*$ & 0.001 $-$ 3 & - \\
Disk inner radius, R$_{\rm in}$      & 0.1 $-$ 5          & AU \\
Disk outer radius, R$_{\rm out}$     & 100 $-$ 500        & AU \\
Disk surface density distribution power, $\gamma$ & 0.5 $-$ 1.5        & -  \\
Disk vertical scale height, H$_0$    &  3 $-$ 20          & AU \\
Flaring index, $\phi$                & 1 $-$ 2.5          & -\\
Vertical viscosity, $\alpha_v$       & 10$^{-5}$ $-$ 0.1  & - \\
Dust size power index, $\beta$       & 3.5 $-$ 4.5        & - \\
smallest dust size, a$_{\rm min}$    & 10$^{-3}$ $-$ 5 $\times$ 10$^{-2}$   & $\mu$m \\ 
largest dust size, a$_{\rm max}$     & fixed, 3000        & $\mu$m \\ 
Disk resolution, N$_x$, N$_y$        & fixed, 50, 50      & pixel \\
PAH abundance, X$_{\rm PAH}$         & fixed, 0.01        & -     \\
Distance from Sun, D                 & fixed, 140         & pc    \\ 
Outputs of time-dependent chemistry   & fixed, 0.1, 1, and 3   & Myr    
\enddata
\end{deluxetable*}
 
\begin{figure*}[tb]
\centering
\includegraphics[angle=0,scale=0.7]{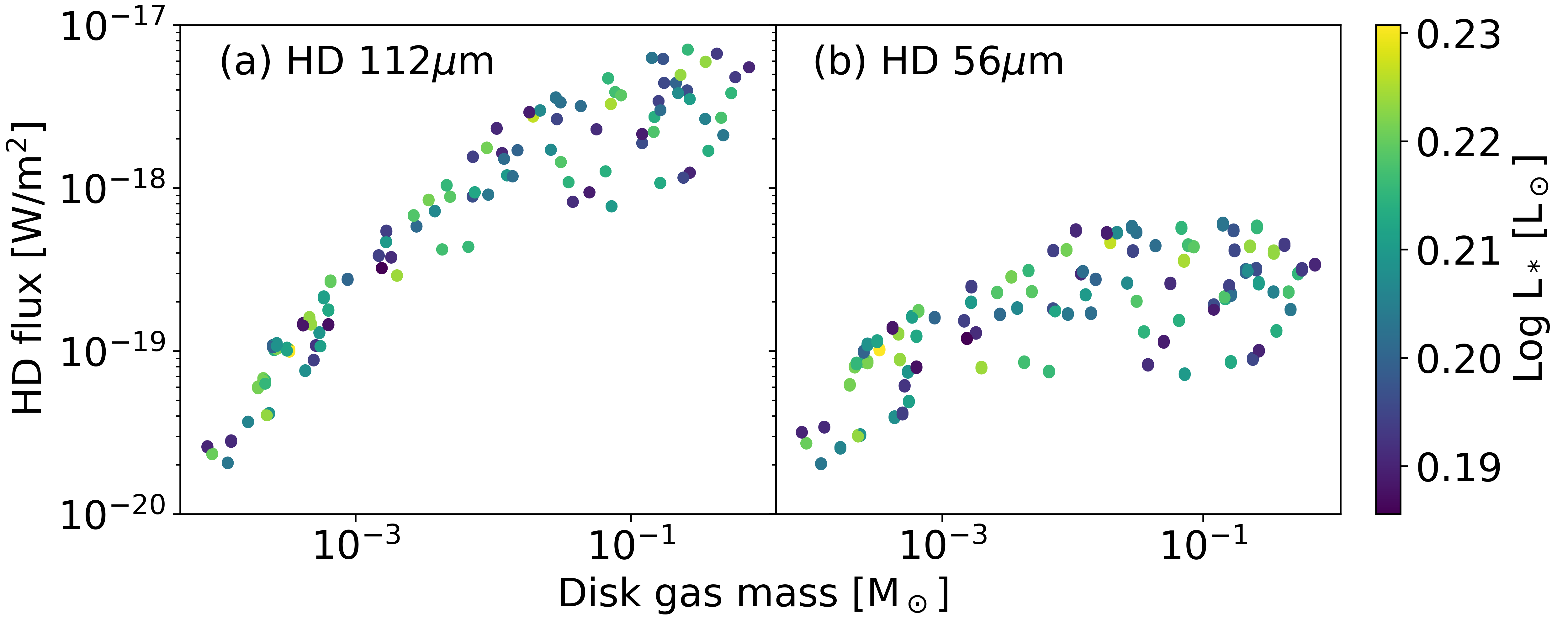}
\caption{Distribution of disk gas mass {\it vs.} HD flux at 112 $\mu$m (left) and 56 $\mu$m (right)  obtained by varying only three physical parameters of disks: disk gas mass, vertical scale height, and flaring index. The colors of the points denote the central star luminosity in units of the luminosity of the sun. The stellar and dust grain properties are fixed in these models. The inclination is fixed at 45 degrees, and the distance from the Sun is fixed at 140 pc. These models replicate the correlated HD fluxes at 56 and 112 $\mu$m from \citet{trapman17}, validating the ProDiMo code.}
\label{f1}
\end{figure*}

To test the chemical network of ProDiMo regarding HD, we first modeled the disks using the same parameter ranges explored by \citet{trapman17} with minor differences in dust properties and radiative transfer. X-ray ionization is not included in our models, while the ISRF is implemented with 10\% of the typical solar neighborhood value. The { disk inclination} is fixed at 45 degrees, while \citet{trapman17} consider disks at 6 degrees. Only the disk mass, scale height, and flaring index are varied. Figure \ref{f1} shows a monotonic relationship between the HD 112 $\mu$m flux and the disk gas mass, similar to Figure 2 in \citet{trapman17}. The HD fluxes from our models are slightly lower (by approximately 25\%) due to different dust optical properties and disk inclinations, which may increase the { overall disk} opacity { including HD line opacity}. However, this confirms that the configurations of the ProDiMo code in this study successfully reproduce the correlation between the HD flux and the disk gas mass as shown in \citet{trapman17} and that HD fluxes are a good tracer for measuring the disk gas mass if the disk properties are very well constrained.   

The focus of this study is to explore the relationship between physical parameters and observables beyond the parameter ranges considered { in} previous studies and to find the optimal set of measurements to constrain the disk gas mass. To probe the statistical relationship more accurately, we use a Monte Carlo approach. Monte Carlo simulations are often used to explore output distributions as a function of the input distributions when the { dependency} between the input and output quantities is complicated. The same method has been applied { in} different studies and demonstrated to facilitate science retrievals with the corresponding uncertainties \citep[e.g.,][]{seo23}. In this work, we may view the physical parameter as input. Physical, chemical, and radiative processes are the functions/operators transforming the input distributions into the output distributions, while the distribution of observables is the output distribution. We assume 15 free disk parameters during the Monte Carlo modeling and use the ProDiMo code to { compute} the observables. 

There are more parameters that can be adjusted in the code, but the 15 parameters already make the modeling a high-dimensional problem. The distribution of observables can be expressed as a function of physical parameters or vice versa. We do not use a grid of models to probe the prior and posterior space since the discrete selection of parameters appears as an artificial pattern in the input distribution, which may be passed down to the output distribution, increasing noise in the output distribution. 

\begin{figure*}[tb]
\centering
\includegraphics[angle=0,scale=0.8]{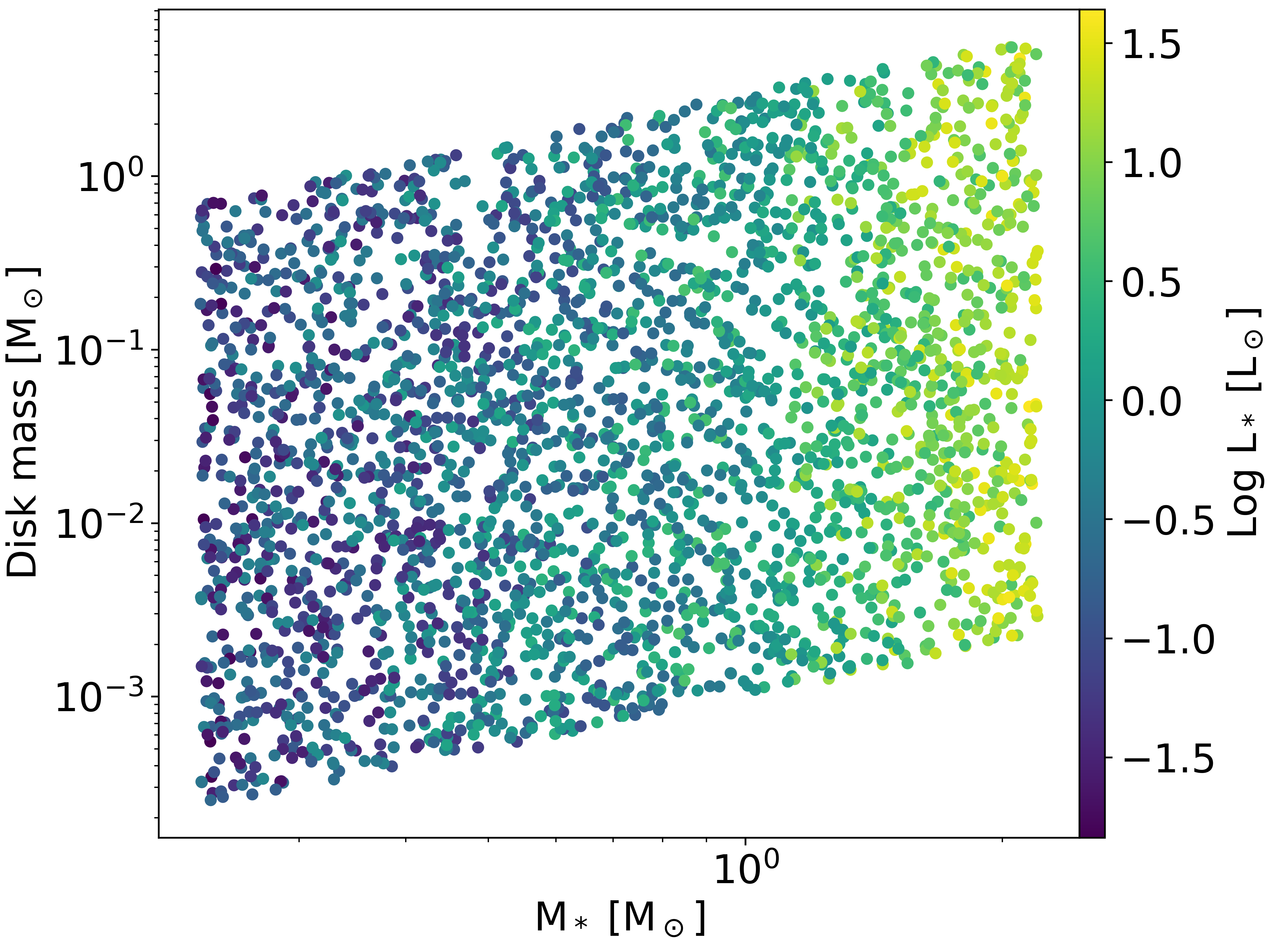}
\caption{The distribution of 3000 models in the central star mass {\it vs.} disk total mass space. The colors of the points in the scatter diagram denote the central star luminosity in units of the luminosity of the sun. Note that the disk total mass is not a free parameter, but the ratio of disk mass to the central star mass is a free parameter. See Table \ref{tab:para} for the free parameters and their ranges.}
\label{f2}
\end{figure*}

Table \ref{tab:para} shows the 15 free parameters for the Monte Carlo method and their corresponding ranges. Most of the parameters are selected randomly with uniform distributions on a log scale. We work in log space to cover the wide dynamic range of the physical parameters and analyze the statistical relationships in log-log space. We use uniform distributions on a log scale since many physical parameters follow the power-law distribution (e.g., stellar mass, \citet{lada03}). There are a few parameters that we do not sample in the log space. These parameters either have a narrow range (disk scale height) or are exponent values (disk surface density power index, dust size distribution power index, disk flaring index). A random selection in the uniform distributions is assumed since we do not know the distributions of these parameters in reality. If the distributions of any of the parameters are known, we can resample from the current distribution to match the known distributions. Figure \ref{f2} shows the distribution of the disk mass {\it vs.} stellar mass for the 3000 disk models, with the color of points denoting the luminosity of the central star. The range of physical parameters in this study is significantly larger than the physical parameters found in observed disks (e.g., $M_{\rm disk} > 0.1 M_\odot$). We intentionally explore a wide range of physical parameters beyond observed values to { isolate retrieval} uncertainties {originating from the thermochemical models since the uncertainty ranges may go beyond the observed parameter ranges. If the range is limited by the observed value, the retrieval uncertainty cannot be fully measured, and it is often underestimated by the limitation of parameter ranges.} { On the other hand, the models having parameters beyond the observed ranges may not describe realistic disks. For example, a disk with a mass significantly larger than 0.1 M$_\odot$ is likely to have an envelope surrounding the system, which we do not consider in the model. Thus, the retrieval of disk gas mass beyond 0.1 M$_\odot$ using these 3000 models should not be taken as an accurate solution. }

{ The number of the models is determined by the target uncertainty of the retrieval. We use a statistical approach for science retrieval works in a two-dimensional space (e.g., HD flux and disk gas mass), and we target a resolution (average spacing between data points of models) in a single dimension/axis to be less than 0.17 on a log scale, which is a spacing equivalent to 50\% of the data point value. In this study, the average spacing between the data points of models in disk gas mass is about 0.1 in log space, which is a sufficient resolution for this study. We further elaborate on the resolution and its implication to the statistical results in \S\ref{sec:stats_1}.  }

\begin{figure*}[tb]
\centering
\includegraphics[angle=0,scale=0.8]{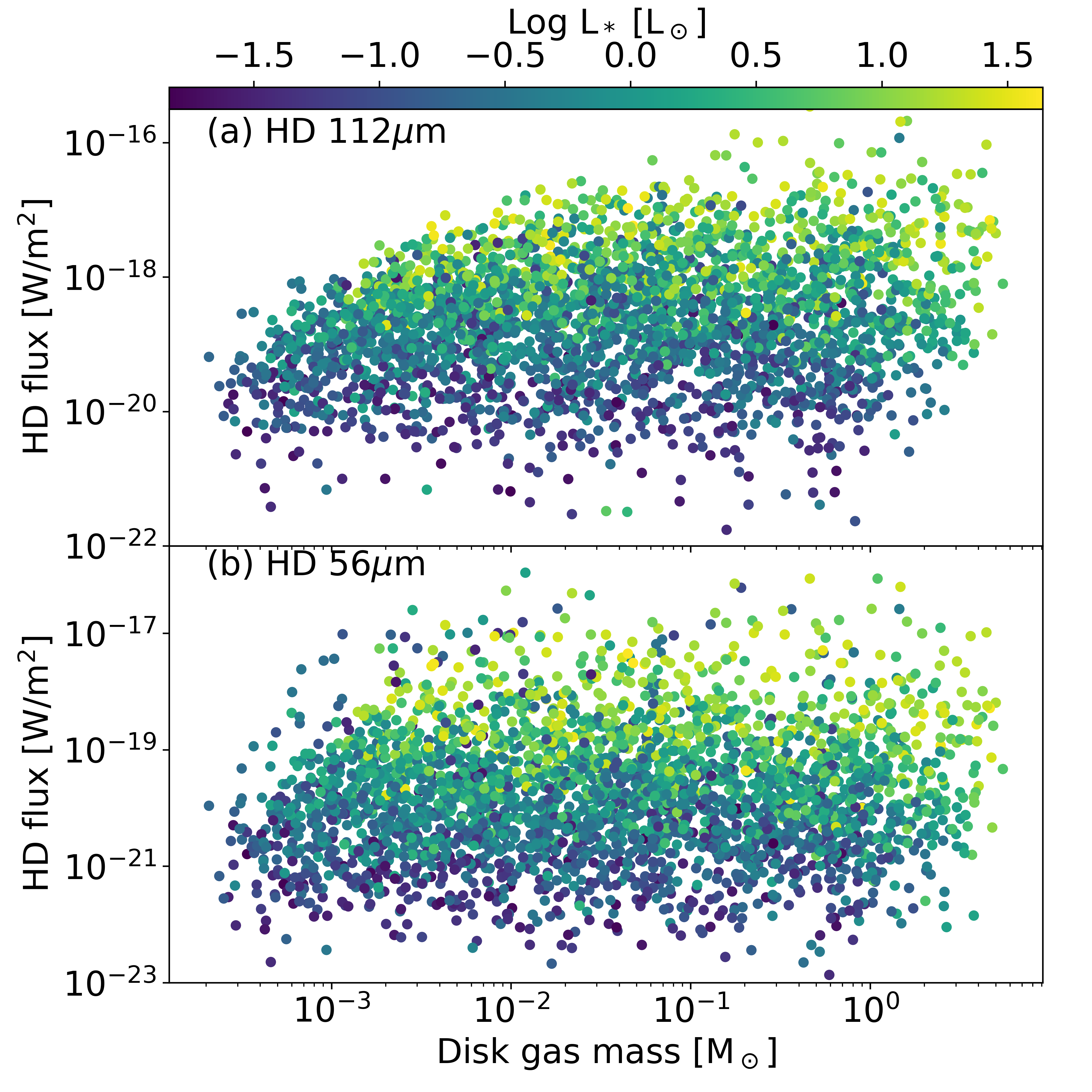}
\caption{The distribution of the HD fluxes at 112 $\mu$m (top) and 56 $\mu$m (bottom) as a function of the disk gas mass. The colors of the scatter points denote the central star luminosity in units of the luminosity of the sun. The models have 15 free parameters. The inclination and the distance are fixed to be 45 degrees and 140 pc, respectively. The chemical age of the disk is 1 Myrs; however, we found that the HD fluxes vary only minimally ($<$10\%) over the modeling time frame (0 to 3 Myr). When considering the full parameter space, the models scatter significantly and do not show a tight correlation between the HD fluxes and the disk gas mass. However, the colors of the points show a correlation between the central star luminosity and the HD flux, suggesting that the central star luminosity is a key parameter driving the HD flux. }
\label{f3}
\end{figure*}

\section{HD Line Flux and Disk Gas Mass} \label{sec:HD}

Figure \ref{f3} shows the 56 $\mu$m and 112 $\mu$m HD line fluxes at 1 Myr as a function of the disk gas mass for the full model range. We made outputs at 0.1, 1, and 3 Myr, but we only show the results at 1 Myr since the HD fluxes vary minimally ($<$ 10\%). The disk inclination is fixed at 45 degrees, and the distance is assumed to be 140 pc from the Sun. The figure shows that the HD flux for a given disk gas mass has an upper limit. The 56 $\mu$m and 112 $\mu$m fluxes are spread widely below this limit. The trend is similar for both the 56 $\mu$m and 112 $\mu$m fluxes, while the detailed shapes of the upper limits differ. The upper envelope of the 56 $\mu$m fluxes flattens for disk gas masses higher than a few times 10$^{-3}$ M$_\odot$. The upper envelope of the 112 $\mu$m fluxes increases slowly when the disk gas mass becomes higher than 10$^{-2}$ M$_\odot$. This trend is similar to the monotonic correlation between the HD flux and the disk gas mass shown in \citet{trapman17}. The overall distribution of the HD fluxes does not show any strong correlation with the disk gas mass when all disk parameters are varied freely, demonstrating the need for the addition of known priors, such as stellar luminosity. 

\begin{figure*}[tb]
\centering
\includegraphics[angle=0,scale=0.7]{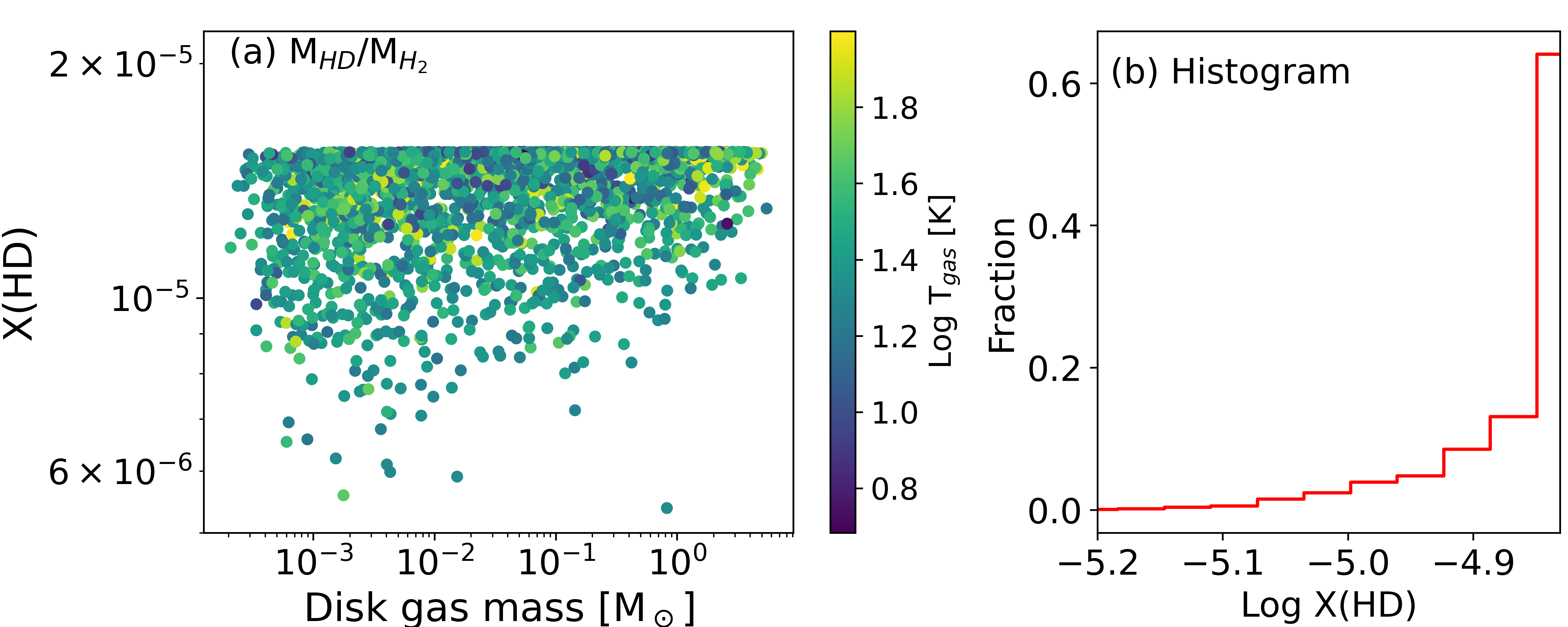}
\caption{HD fractional abundance as a function of disk gas mass (left) and the histogram of the HD fractional abundance (right). The colors of the points in the scatter diagram denote the mass-weighted average gas temperatures of the disks. The HD fractional abundance is mostly constant with small variations of less than a factor of two, agreeing with the results of \citet{trapman17} and \citet{kama20}. The variation becomes larger when the disk gas mass is smaller. The optical depth decreases with mass, and the UV photons may penetrate deeper into the disks, dissociating more H$_2$ and HD molecules. A constant HD fractional abundance strongly suggests that the HD factional abundance is not the determining factor of the large variation in the HD fluxes.}
\label{f4}
\end{figure*}

\begin{figure*}[tb]
\centering
\includegraphics[angle=0,scale=0.73]{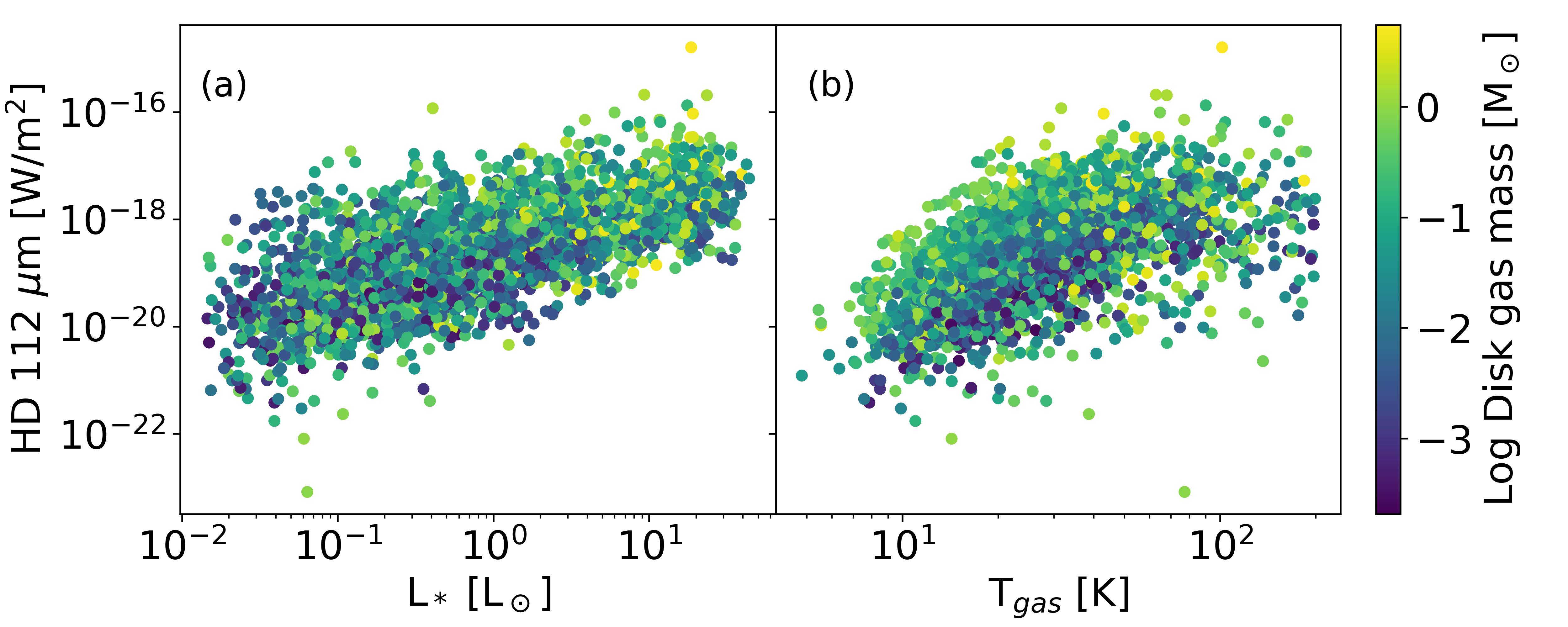}
\caption{HD fluxes at 112 $\mu$m as a function of the central star luminosity (left) and the mass-averaged gas temperature (right). The colors of the points denote the disk gas mass. The HD fluxes have positive correlations with the central star luminosity and the mass-averaged gas temperature. However, the scatter is relatively large, and the width of the scattering corresponds to the range of the disk gas mass of the models. The right panel shows a weak color gradient of the points (bright green to dark green from top to bottom), which suggests that there may be systematic variations of the HD flux. On the other hand, there is no trivial color gradient in the left panel. }
\label{f5}
\end{figure*}

The scatter in Figure \ref{f3} likely originates from the large parameter space that affects the HD molecules' excitation condition rather than from HD's chemistry. Figure \ref{f4} (a) shows the overall fractional abundance of HD as a function of the disk gas mass at 1 Myr. Results at 0.1 and 3 Myr show that the HD fractional abundance varies by less than 5\% over these times (mostly around 1\%). The HD fractional abundance is relatively constant as a function of the disk gas mass, with only small variations compared to the HD flux (Figure \ref{f4} (b)). Over 90\% of the 3000 disk models have fractional abundances between 1 $\times$ 10$^{-5}$ and 1.3 $\times$ 10$^{-5}$. The small variation (less than a factor of 2) in the HD fractional abundance is because H$_2$ molecules efficiently shield the UV photons, preventing HD from dissociating, even for the models with large variations in the UV field (five orders of magnitude). The small variation in x(HD) is mainly due to a minor increase in the penetration of UV photons through the disk with a larger input of UV flux from the central star and the ISRF. Also, the low-mass disks, having lower optical depths, tend to show larger scatter in the HD fractional abundance. The small variation in { X}(HD) suggests that the HD abundance is not the main reason for the large variations present in the HD fluxes. 

Figure \ref{f5} shows the 112 $\mu$m HD flux as a function of the luminosity of the central star (panel (a)) and the mass-averaged disk gas temperature (panel (b)). The HD flux is positively correlated with both the stellar luminosity and the gas temperature. This suggests that the HD flux is connected to the radiative transfer process since the stellar luminosity and the gas temperature determine the excitation conditions of the HD molecules. There are roughly four orders of magnitude variations in the HD flux for a given stellar luminosity or a given disk gas temperature. These variations are comparable to the range of the disk gas mass. 

\begin{figure*}[tb]
\centering
\includegraphics[angle=0,scale=0.7]{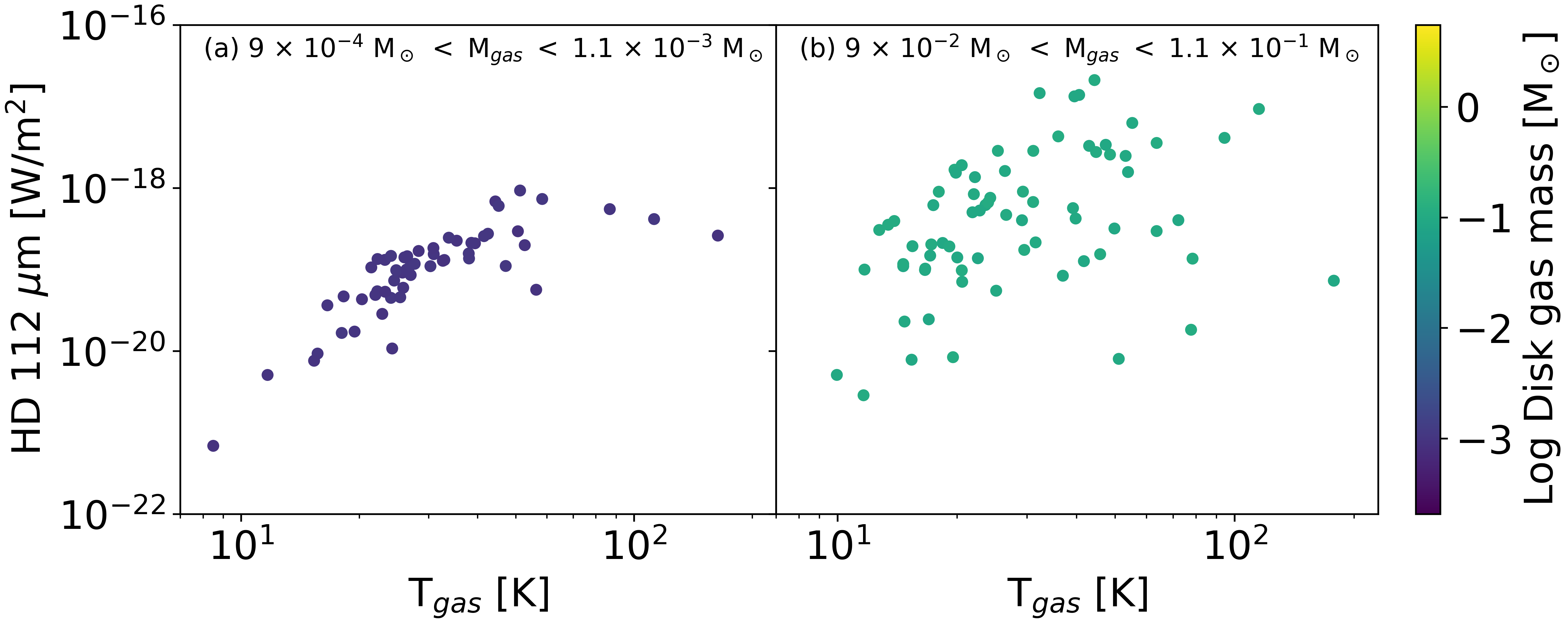}
\caption{HD flux at 112 $\mu$m over a limited range of disk gas mass as a function of the mass-averaged gas temperature.  Panel (a) represents low-mass, optically thin disks, while  
panel (b) shows massive, optically thick disks. The HD fluxes have a strong correlation with the mass-averaged gas temperature when the disks are optically thin. On the other hand, the HD fluxes show a large scatter for optically thick disks. This suggests that the excitation condition of the HD molecules and the optical depth determine the HD flux.  }
\label{f6}
\end{figure*}
\begin{figure*}[tb]
\centering
\includegraphics[angle=0,scale=0.7]{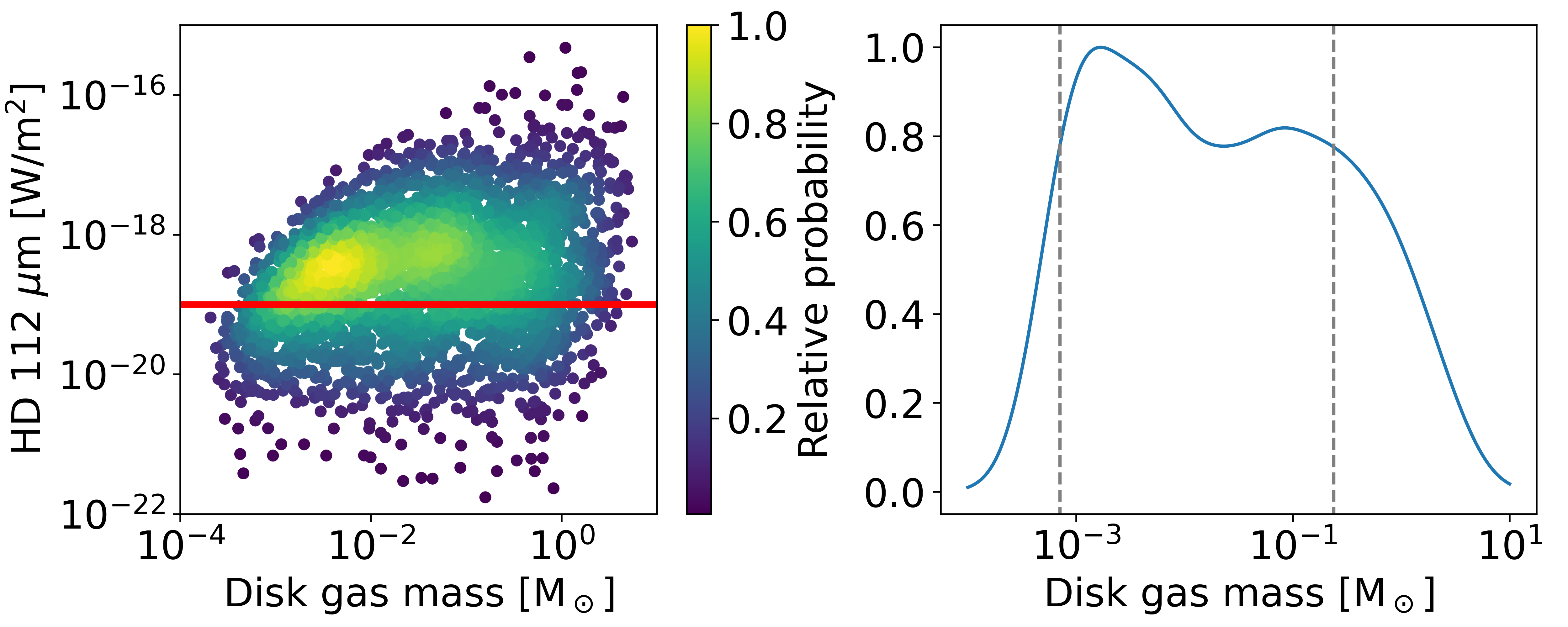}
\caption{(left) HD flux at 112 $\mu$m as a function of disk gas mass. The colors of the points denote the 2-dimensional relative probability density function (PDF) estimated from the density of the scatter points using the Gaussian Kernel Density Estimation (KDE). The kernel size of the Gaussian KDE is determined to have at least six points on average. (right) Relative probability of the disk gas mass when the HD flux is measured to be 10$^{-19}$ W m$^{-2}$ (solid red line in the left panel), which is a slice through the 2-dimensional PDF. The dashed gray lines denote the 1-$\sigma$ uncertainty range of the disk gas mass. This figure shows that the disk gas mass cannot be accurately determined when the HD flux at 112 $\mu$m is the only information known.}
\label{f7}
\end{figure*}

Figure \ref{f6} shows the HD flux at 112 $\mu$m within two different ranges of the disk gas mass in order to assess the sensitivity of the HD flux to the disk gas temperature { (see Appendix for discussion of the HD flux dependence on other physical parameters.)}. Panel (a) shows the case for a low-mass ($\sim$10$^{-4}$ M$_\odot$) disk { when HD 112 $\mu$m line may be optically thin}. The HD 112 $\mu$m flux increases monotonically with the gas temperature up to 50 $-$ 60 K and seems to saturate or slightly decrease at a higher temperature. The saturation is due to the upper energy state of the HD 1 $-$ 0 transition being $E/k$ = 128.4 K. At a higher temperature, the population of HD at the first excited state reduces, lowering the frequency of HD 1 $-$ 0 transition. This suggests that the gas temperature mainly determines the HD flux at 112 $\mu$m if the disk is optically thin. On the other hand, optically thick disks display different behavior, showing a large scatter in the HD fluxes. Panel (b) shows the 112 $\mu$m HD flux for massive disks ($\sim$10$^{-1}$ M$_\odot$). In this mass range, the  112 $\mu$m HD line is optically thick, and the HD flux is determined by the gas temperature at the photo surface where the optical depth of HD is unity. The location of the photo surface is determined by the disk's physical structure. Since we cover a large range of physical structures, { a} large scatter in panel (b) is understandable.

\section{Statistical Prediction of Disk Gas Mass} \label{sec:stats}
\subsection{Statistics and probability density function}\label{sec:stats_1}

The details of the statistical approach used in our study can be found in \citet{seo23}. Here, we briefly describe the method. The main goal of this study is to find the most probable disk gas mass for a given set of known observables{/information} (such as line emissions, { dust} SED or other physical parameters, for example, central star mass, disk size, etc.) and the uncertainty of the most probable disk gas mass estimate. The most probable output and its uncertainty can be directly estimated if the probability density function (PDF) of the output quantities as a function of input parameters is measured. A popular approach to estimate the PDF is to use Monte Carlo sampling of output values for randomly selected input values, make a scatter plot in the input {\it vs.} output space, and measure the density of the scatter points \citep[e.g.][]{rosenblatt56,parzen62,Kroese14}. The density of the scatter points is equivalent to the occurrence rate of the points; therefore, the density function/field of the scatter points is the PDF of the output values as a function of input values or vice versa.

The density measurement of scattered points is estimated using a Gaussian kernel density estimation (Gaussian KDE) from the scikit-learn package of Python scikit-learn \citep{pedregosa11}. We normalize the density peak to unity. We use relative probability rather than absolute probability since we focus on deriving the most probable values of the disk gas mass and its uncertainty for a given set of observables, which is the same either using relative or absolute probability. The kernel width is fixed to have six points/models on average within the kernel full-width half maximum (FWHM); thus, it depends on $\sim\sqrt{N}$, where $N$ is the total number of models. The width of a PDF containing 68\% of the total area within the PDF represents the 1-$\sigma$ uncertainty, so the uncertainty in this study cannot be smaller than the kernel width. With 3000 models, the average spacing between scatter points is typically about 0.1 in log space, which is a factor of $\sim$ 1.25 and is the minimum uncertainty resolution.  

\subsection{Most probable disk gas mass and uncertainties}\label{sec:stats_2}
\subsubsection{Constraining disk gas mass using only HD flux}

The left panel of Figure \ref{f7} shows the same distribution of 112 $\mu$m HD fluxes as a function of the disk gas mass as Figure \ref{f3}, except that here the color of data points denotes the relative probability. The highest probability for a given disk gas mass is typically between 10$^{-19}$ W m$^{-2}$ and 10$^{-18}$ W m$^{-2}$, suggesting that a telescope sensitivity in this range may be { required} for surveying the HD 112 $\mu$m emission from protoplanetary disks at distances of $<$140 pc. The right panel of Figure \ref{f7} shows the relative probability when the HD flux at 112 $\mu$m is measured to be 10$^{-19}$ W m$^{-2}$ but without any other information being known about the disk (a cut along the red line in the left panel). The most probable disk gas mass, based on the location of the peak relative probability, is 1.3 $\times$ 10$^{-3}$ M$_\odot$, but the uncertainty of this value is very significant. The 1-$\sigma$ uncertainty is the range of the disk gas mass where the integrated PDF within the range becomes 68\% of the total integration of the PDF. The measured 1-$\sigma$ uncertainty here is nearly three orders of magnitude, and there is also the second probability peak within the 1-$\sigma$ uncertainty range. This confirms that the HD flux at 112 $\mu$m is not on its own a good tracer for measuring the disk gas mass. However, with high-resolution images from large telescopes, a few physical parameters of the disks and their environment (e.g., central star luminosity) are typically known. It is important to know which set of parameters are key to determine the disk gas mass.

\subsubsection{Constraining disk gas mass using HD flux and physical parameters}\label{sec:stats_3}

We probe the minimum information required to estimate the disk gas mass accurately using the HD flux. \citet{trapman17} clearly demonstrated that the HD flux is well-correlated with the disk gas mass when there is abundant information about the disk and the central star. However, the physical parameters that are fixed in their models (e.g., dust size distribution, cosmic-ray flux, central star luminosity, etc.) are barely determined for many observed disks. Such physical parameters are not easy to determine from observations and often have large uncertainties. Here, we estimate how the PDFs vary with the number of known physical parameters and determine the uncertainty in the most probable disk gas mass as a function of the number of known physical parameters.  

To explore how additional measurements can be used in conjunction with the HD flux to constrain the disk gas mass we consider the case of Model 414 (Table~\ref{tab:414}) and adopt its physical values as the true solution, adding Gaussian uncertainties as follows: the HD flux at 112 $\mu$m is 2.87$^{+2.87}_{-1.44}$ $\times$ 10$^{-18}$ W m$^{-2}$, the central star luminosity is 1.28$^{+0.33}_{-0.26}$ L$_\odot$, the disk size is 400 $\pm$ 20 AU, the power index of dust size distribution is 3.83 $\pm$ 0.2, and the minimum dust size is 2.0$^{+0.2}_{-0.1}$ $\times$ 10$^{-3}$ $\mu$m. The uncertainties in the HD flux, the central star luminosity, and the minimum size of dust grain are asymmetric since they are Gaussian on a log scale. The uncertainty of the HD flux is assumed to be relatively larger (a factor of 2) than typical absolute flux uncertainties (a factor of $<$1.5), so we have a reasonable number of models within the HD uncertainty range to convolved with the uncertainty ranges of other parameters. We use Model 414 as a toy model because it is a massive disk (an optically thick disk). Such a massive disk is rare in reality \citep{Ansdell16}, but the uncertainty of disk gas mass estimation of an optically thick disk is larger than that of optically thin disks. Thus, the uncertainty of disk gas mass evaluated using Model 414 will serve as the upper limit of the uncertainty in estimating disk gas mass of optically thin disks.     

\begin{figure*}[tb]
\centering
\includegraphics[angle=0,scale=0.7]{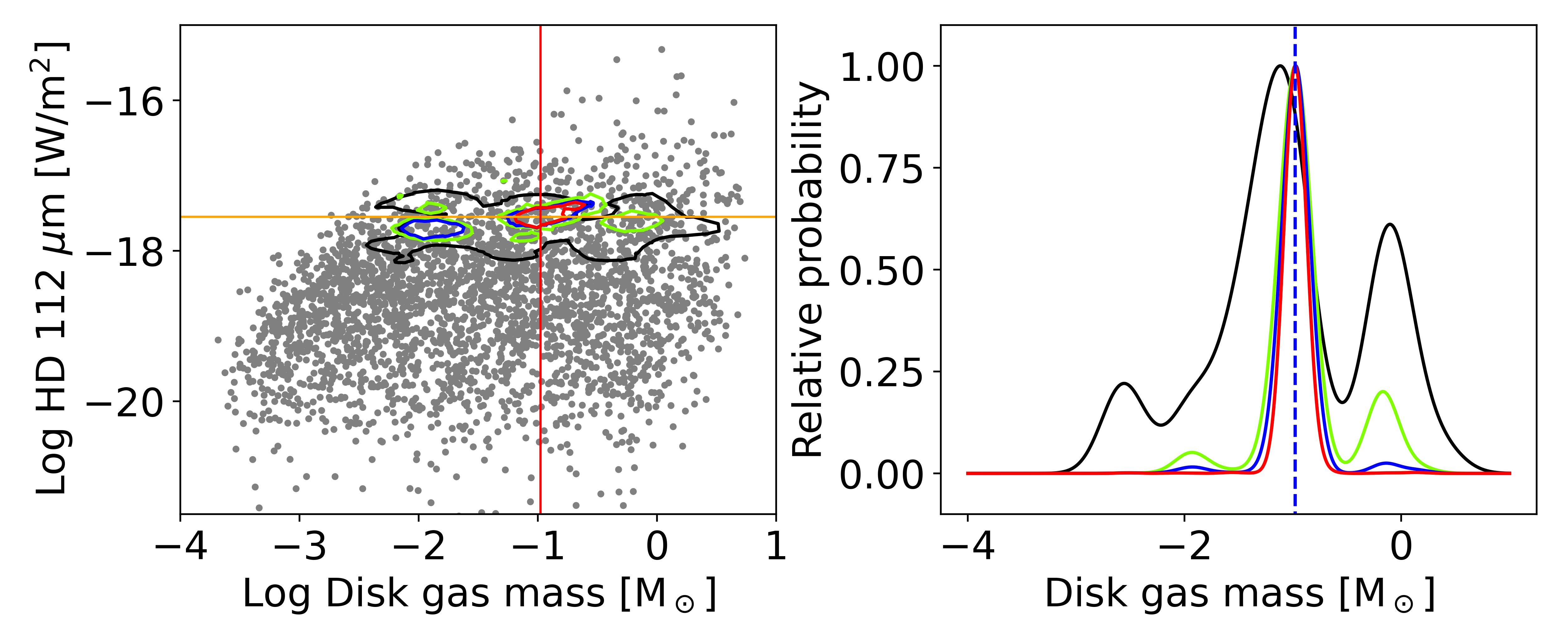}
\caption{(left) HD flux at 112$\mu$m as a function of disk gas mass (left). The contours denote the 1-$\sigma$ uncertainties of PDFs for different sets of known physical parameters (summarized in Table \ref{tab:set}). The black contour (set A) applies when the HD flux and the central star luminosity are known with given uncertainties. The green contour (set B) applies when the HD flux, the central star luminosity, and the disk size are known. The blue contour (set C) applies when the power index of dust size distribution is additionally known to the green contour case. Finally, the red contour (set D) represents the case when the minimum dust size is determined in addition to the four parameters of the blue contour case. The red vertical line denotes the true solution of this exercise. (right) Most probable disk masses { are} based on the HD flux (orange horizontal line in the left panel).  The color of the lines corresponds to the same cases as the contour lines in the left panel. The relative PDFs show a decrease in the uncertainty as more physical parameters are known. We found that the uncertainty is adequately small when at least four physical parameters are known (red line).   }
\label{f8}
\end{figure*}

Figure \ref{f8} shows the relationship between the HD 112 $\mu$m flux and the disk mass (left){ ,} with contours showing the uncertainties in the mass determined for each of the parameter sets considered. The orange horizontal line denotes the observed HD flux, and the red vertical line marks the true solution of the disk gas mass for this exercise. Figure \ref{f8} (right) shows the PDFs for the same sets of parameters, estimated using the Gaussian KDE with weights applied to each model based on the uncertainty of each physical parameter. Here{,} the blue vertical dashed line is the true solution { to} the disk mass. We consider four sets of parameters summarized in Table \ref{tab:set}: (A) HD flux at 112 $\mu$m and the central star luminosity (shown in black). (B) To the parameters from (A) we add the disk size (green), (C) To the parameters from (B) we add the power-law index of the dust size distribution (blue), and (D) to the parameters form (C) we add the minimum dust size (red). The green and blue contours have multiple peaks since they have multiple local maxima in the PDFs (see the right panel). 
\begin{deluxetable*}{lll}
\tablecaption{\label{tab:set}Parameter sets used to determine uncertainties in disk masses in Figure \ref{f8} and the uncertainties of the disk gas mass.}
\tablewidth{0pt}
\tablehead{
\colhead{Label} & \colhead{Parameters} & \colhead{1-$\sigma$ of disk gas mass} 
}
\startdata
A & HD flux at 112$\mu$m and L$_*$  & 0.023 $-$ 1.33 M$_\odot$\\
B & HD flux at 112$\mu$m, L$_*$, and R$_{\rm out}$ & { 0.061 $-$ 0.17 M$_\odot$}\\
C & HD flux at 112$\mu$m, L$_*$, R$_{\rm out}$, and $\beta$ &  { 0.077 $-$ 0.14 M$_\odot$}\\
D & HD flux at 112$\mu$m, L$_*$, R$_{\rm out}$, $\beta$, and a$_{\rm min}$ & { 0.082 $-$ 0.13 M$_\odot$} \\
\enddata
\end{deluxetable*}

The 1-$\sigma$ uncertainty contours reduce as we add more information about the disk and the central star, suggesting that the uncertainty in the disk gas mass measurement will be significantly reduced as we know more information about the disk. We also considered other physical parameters but found, through trial and error, that { the HD flux, the central star luminosity, and the disk size are the main contributors containing the disk gas mass. As for the next most significant information, we found that the dust parameters provide the best reduction in uncertainty, while the disk structure parameters (e.g., disk scale height, disk inner rim radius) also make a minor reduction in uncertainty.} 

In the PDF plot, the most probable disk mass indicated by parameter set A (shown in black) is slightly off from the true solution, while its range (the PDF width) is 0.023 $-$ 1.33 M$_\odot$, which is about a factor of 50. The most probable values of the green (B), blue (C), and red (D) PDFs align well with the true solution. However, the uncertainty is the smallest (0.082 $-$ 0.13 M$_\odot$) for the red PDF, which has four physical parameters known along with the HD flux. The 1-$\sigma$ uncertainty of the green PDF is 0.061 $-$ 0.17 M$_\odot$. The uncertainty of the blue PDFs is 0.077 $-$ 0.14 M$_\odot$. This suggests that the HD flux at 112 $\mu$m may correlate well with and is a good tracer for the disk gas mass when we know the central star luminosity, disk size, and dust size distribution. However, estimating these parameters is often not trivial, and thus, an observational constraint that relates to these parameters may be a good substitute for these parameters in estimating the disk gas mass.  

\begin{deluxetable*}{cccccc}
\tablecaption{Parameters of Five Models\label{tab:414}}
\tablewidth{0pt}
\tablehead{
\colhead{Parameter} & \colhead{Model 414} & \colhead{Model 1251} & \colhead{Model 2346} & \colhead{Model 6033} & \colhead{Model 6042}
}
\startdata
Stellar mass, M$_*$ [M$_\odot$]            & 0.66       & 0.72      & 0.69  & 0.29   & 0.86    \\
Central star luminosity, L$_*$ [L$_\odot$] & 1.28       & 1.07      & 0.36  & 0.41   & 1.50    \\
Central star temperature, L$_*$ [K]        & 4084       & 4125      & 3944  & 3425   & 4299    \\
UV fraction, F$_{\rm UV}$ [10$^{-2}$]      & 0.01       & 0.0003    & 0.08  & 1.06   & 0.0014  \\
Cosmic-ray flux, F$_{\rm CR}$ [10$^{-19}$ cm$^{-2}$] & 0.4  & 0.15  & 0.27  & 6.35   & 2.68    \\
ISRF, F$_{\rm ISM}$ [G$_0$]                & 2.04       & 0.06      & 0.97  & 0.78   & 3.13    \\
dust/gas ratio, $\rho_d$/$\rho_g$          & 0.0130     & 0.029     & 0.013 & 0.016  & 0.089   \\
Disk mass [M$_\odot$]                      & 0.106      & 0.0010    & 0.052 & 0.0053 & 0.011   \\
Disk inner radius, R$_{\rm in}$ [AU]       & 0.50       & 4.65      & 2.3   & 0.13   & 0.99    \\
Disk outer radius, R$_{\rm out}$ [AU]      & 401        & 272       & 184   & 258    & 474     \\
Disk surface density power, $\gamma$       & 1.13       & 0.82      & 1.21  & 1.39   & 1.41    \\
Disk vertical scale height, H$_0$ [AU]     & 14.7       & 18.2      & 10.0  & 16.3   & 17.8    \\
Flaring index, $\phi$                      & 1.67       & 1.25      & 1.65  & 1.46   & 1.31    \\
Vertical viscosity, $\alpha_v$ [10$^{-3}$] & 0.286      & 1.29      & 70    & 0.981  & 0.65    \\
Dust size power index, $\beta$             & 3.84       & 3.59      & 3.50  & 4.13   & 3.84    \\
smallest dust size, a$_{\rm min}$ [10$^{-3}$ $\mu$m] & 2.0  & 26    & 14    & 35     & 3.9     \\ 
\enddata
\end{deluxetable*}

\begin{figure*}[tb]
\centering
\includegraphics[angle=0,scale=0.7]{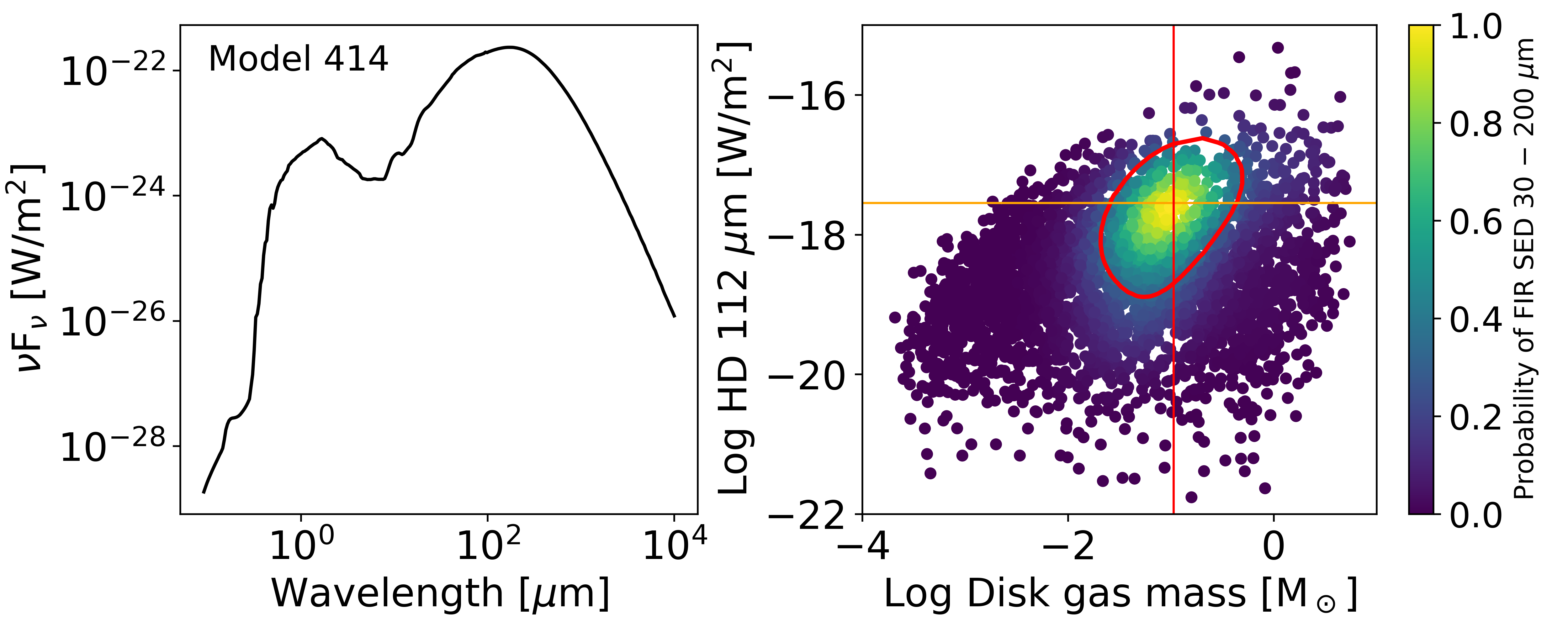}
\caption{Spectral energy distribution of Model 414 (left) and the HD flux at 112 $\mu$m as a function of the disk gas mass with the colors of the scatter points denoting the relative PDF based on the SED fitting (right). Model 414 is assumed to be the observed SED after adding a certain white noise. The red contour is the 1-$\sigma$ uncertainty when the noise-added SED of Model 414 is fitted to the 3000 models. The signal-to-noise ratio of the Model 414 SED is 10 in every channel. The orange and red lines are the true solution of the HD flux and the disk gas mass of Model 414, respectively. The 1-$\sigma$ uncertainty of the disk gas mass using the observed SED and the HD flux is evaluated by the length of the orange line intersecting the red contour. The disks have inclinations of 45 degrees and are at a distance of 140 pc. }
\label{f9}
\end{figure*}

\subsubsection{Constraining disk gas mass using HD flux and SED}\label{sec:stats_4}

Since the SED is directly related to the three physical parameters (central star luminosity, disk size, and dust size distribution), we have tested whether or not it can be used to constrain the disk mass based on the HD emission using Model 414 as an example. The SED determined by ProDiMo covers wavelengths from 912 $\rm \AA$ to 10 mm. The resulting SED for Model 414 is similar to that of a Class I protostar, where the IR emission is brighter than the visible and UV. Model 414 is a massive, flared disk with $\sim$0.1 M$_\odot$, and the SED is evaluated with the inclination of 45 degrees, partially blocking the central star (see Table \ref{tab:414}). 

White noise is added to the SED of Model 414 with various signal-to-noise ratios (SNRs) in each spectral channel{ , as defined below}. The SEDs of all 3000 models (without the addition of white noise) are fitted to the noise-added SED of Model 414 and $\chi^2$ values evaluated for four different wavelength ranges: the UV/blue range from 918 $\rm\AA$ to 2000 $\rm\AA$, visible/near-IR range from 0.4 to 1 $\mu$m, far-IR range from 30 $-$ 200 $\mu$m, and sub-mm/mm range from 0.3 $-$ 6 mm. We resample the SEDs to have the resolving powers of 20, 20, 30, and 50 for UV/blue, visible/near-IR, far-IR, and sub-mm/mm SEDs, respectively.

The $\chi^2$ value indicates the goodness of { the} fit but does not provide the probability. To determine the probability of the disk gas mass in conjunction with the $\chi^2$ value, we need to determine a weight that quantifies the goodness of the SED fit of each model within 0 $-$ 1 and estimate the PDFs using the Gaussian KDE. This is done by using the cumulative $\chi^2$ distribution function instead of the $\chi^2$ distribution function. The $\chi^2$ distribution function gives the probability of a particular $\chi^2$ value. This probability shows how frequently the particular $\chi^2$ occurs. The weight for the Gaussian KDE is the rank of the given $\chi^2$ value convolved with the probability of each $\chi^2$ value so that the perfect fit will have the weight of 1 while $\chi^2$ = $\infty$ returns 0. This can be estimated using the cumulative $\chi^2$ distribution function since it returns the cumulative probability of the $\chi^2$ distribution from $\chi^2$ = 0 (perfect fitting but over-fitting) to a given $\chi^2$ value. This function returns 0 when $\chi^2$ = 0 and 1 when $\chi^2$ = $\infty$. Therefore, we give the weight of each model from the SED fitting as 1 - CDF($\chi^2$, D$_f$), where CDF is the cumulative $\chi^2$ distribution function, and D$_f$ is the number of degrees of freedom. The right panel of Figure \ref{f9} shows the HD flux as a function of the disk gas mass, with the colors representing the PDF of the far-IR SED fitting.

\begin{figure*}[tb]
\centering
\includegraphics[angle=0,scale=0.8]{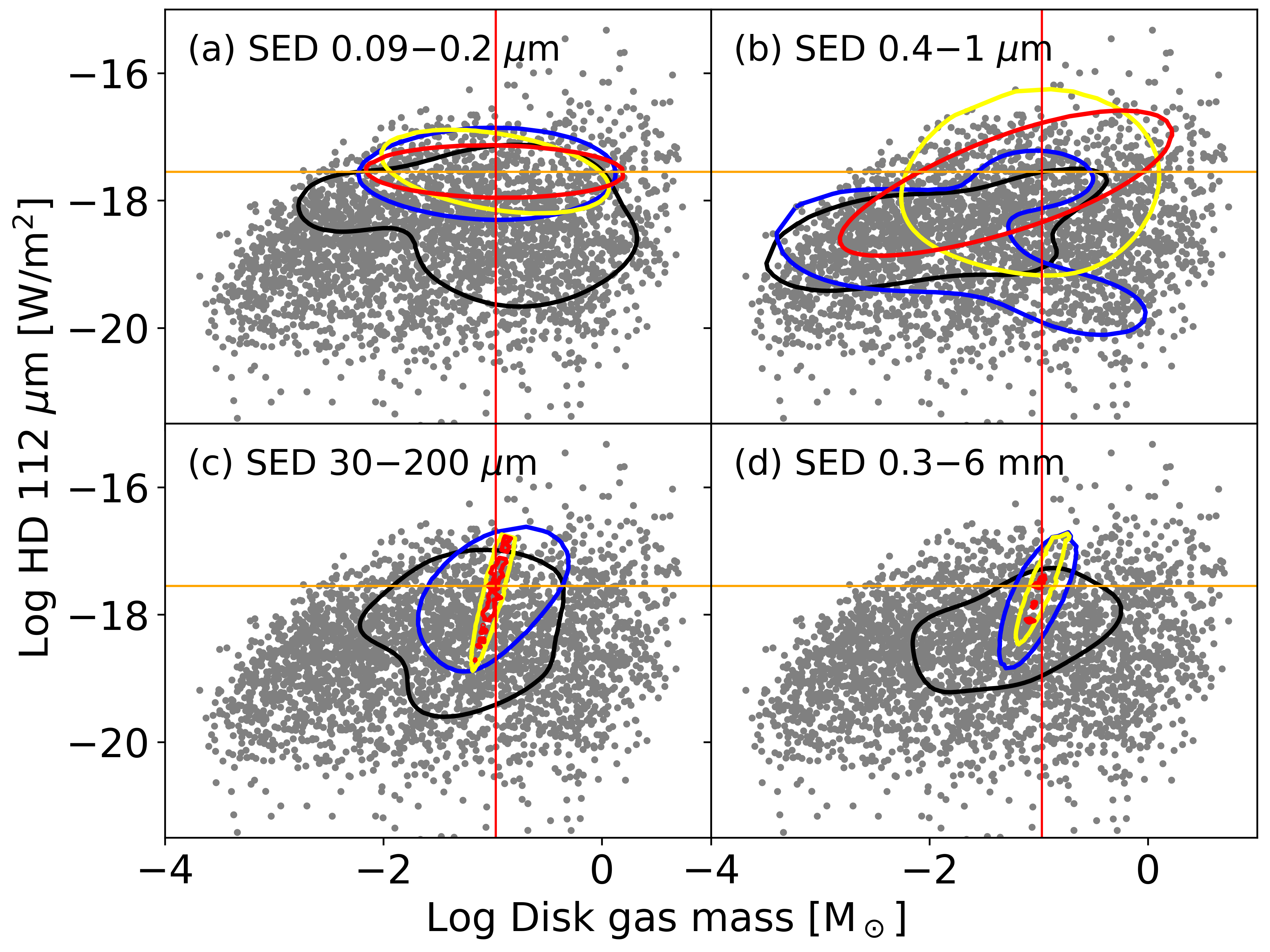}
\caption{HD flux as a function of the disk gas mass with the 1-$\sigma$ uncertainty contours of the SED fitting. The wavelength ranges of the SED fitting are noted in each panel. The different contour colors refer to different SNR of the SED in every channel: 5 (black), 10 (blue), 15 (yellow), and 20 (red). We used model 414 as a toy model and imposed its SED as observed by adding noise at different SNRs. The 1-$\sigma$ uncertainties of the disk gas mass, when both the HD flux (the orange horizontal line) and SED are observed, are evaluated by the length of the orange line intersecting the contours. The red vertical lines indicate the true solution of the disk gas mass. The SEDs in far-IR and sub-mm/mm are relatively orthogonal to the HD flux measurement when the SNR $>$ 15, resulting in the smallest uncertainties in the estimates of the disk gas mass. }
\label{f10}
\end{figure*}

\begin{figure*}[tb]
\centering
\includegraphics[angle=0,scale=0.8]{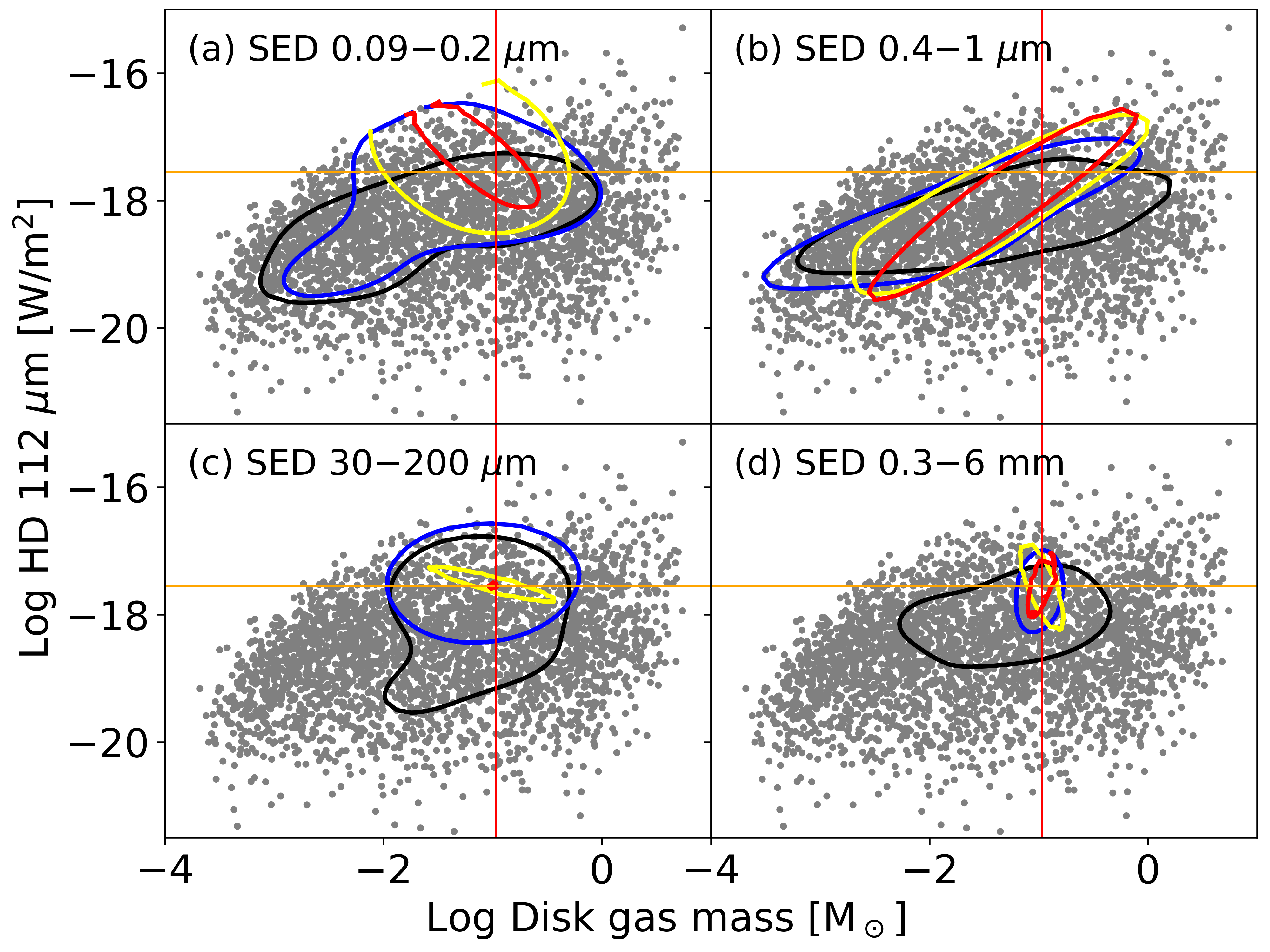}
\caption{Same as Figure \ref{f10} except the inclinations of the disks are 0 degrees.}
\label{f11}
\end{figure*}

Figures \ref{f10} and \ref{f11} show the HD fluxes as a function of the disk gas mass with contours indicating the 1-$\sigma$ uncertainty of the SED fitting at disk inclinations of 45 degrees and 0 degrees, respectively. The black, blue, yellow, and red contours show SNRs of the SEDs of 5, 10, 15, and 20, respectively. The SNR is assumed to be constant throughout every channel. This may not be realistic since the sensitivity of a spectrometer across wavelengths may vary with the spectrometer design, and the SED varies considerably with wavelength. Therefore, the SNR in this study may be taken as an average of SNR across the channels. The orange horizontal line denotes the observed 112 $\mu$m HD flux. The uncertainty of measuring the disk gas mass using both SED and the 112 $\mu$m HD flux refers to the intersection between the orange horizontal line and the contour. Thus, a narrow contour in the horizontal direction suggests a smaller uncertainty in measuring the disk gas mass. The visible/near-IR SED has the largest uncertainty among the four wavelength ranges. The UV/blue SED is also unable to provide a good constraint for the disk gas mass. This is expected since the UV/blue and visible/near-IR SEDs have considerable contributions from scattered stellar light. In section \S\ref{sec:stats_3}, we showed that knowing the central star luminosity does not constrain the disk gas mass at any significant level. 

On the other hand, the far-IR and the sub-mm/mm SEDs show much smaller uncertainties when the SED SNR is larger than 15. The 1-$\sigma$ uncertainty contour is almost orthogonal to the HD flux, reducing the uncertainties of the disk gas mass significantly when both the HD flux and the SED are used together. The uncertainties in the most probable disk gas mass for a disk with an inclination of 45 degrees using the 112 $\mu$m HD flux and either the far-IR or the sub-mm/mm SED with SNR of 20 are less than a factor of 0.5. When the inclination is 0 degrees, the uncertainties are still very small, but the uncertainty using the sub-mm/mm SED slightly increases, while the uncertainty using the far-IR SED decreases. This is reasonable since the SED is sensitive to the optical depth, particularly at shorter wavelengths, and the inclination considerably changes the optical depth along the line of sight through the disk. Thus, the inclination of the disk should be known to be able to accurately constrain the disk gas mass using the HD flux and the far-IR or sub-mm/mm SED.


\begin{figure*}[tb]
\centering
\includegraphics[angle=0,scale=0.8]{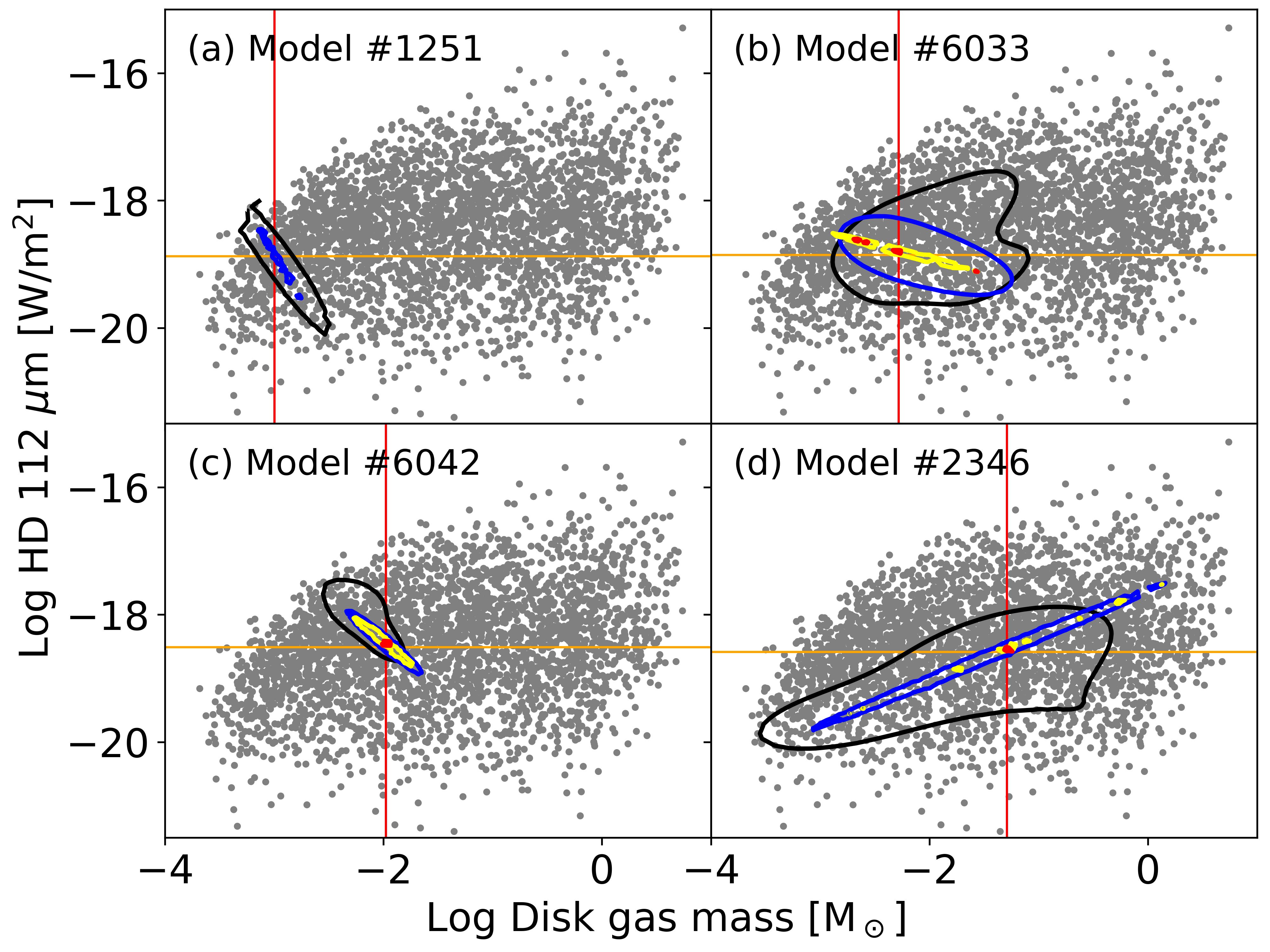}
\caption{HD flux as a function of the disk gas mass with the 1-$\sigma$ uncertainty contours of the far-IR SED fitting. These plots are the same plot of the panel (c) in Figure \ref{f10} but for four different disk cases (see Table \ref{tab:414} for details of the four cases). The number on each panel denotes the disk model number. The different contour colors refer to different SNR of the SED in every channel: 5 (black), 10 (blue), 15 (yellow), and 20 (red). In panel (a), we do not show SNR of 15 and 20 since the uncertainty has already reached its minimum resolution at SNR of 10. The four cases span a wide range of disk gas mass from 5 $\times$ 10$^{-4}$ M$_\odot$ to 0.05 M$_\odot$ to demonstrate that the HD 112 $\mu$m flux and SED may constrain the disk gas mass in a large mass range. }
\label{f12}
\end{figure*}

{ Figure \ref{f12} shows four additional disk gas mass retrieval examples using the HD 112 $\mu$m flux and the far-IR SED. We selected four examples to cover the disk gas mass range from 0.0005 M$_\odot$ to 0.05 M$_\odot$. The five cases, including the case in Figure \ref{f10}, cover a majority of the disk gas mass range anticipated from existing dust mass measurements of protoplanetary disks, assuming a gas-to-dust ratio of 100 \citep{Manara23}. We have not tested the gas mass retrieval for a disk larger than 0.1 M$_\odot$ because our models do not include physics such as envelope and accretion from the envelope, which is often observed toward massive disks with $>$0.1 M$_\odot$.

All of the four cases in Figure \ref{f12} demonstrate that the uncertainty of the disk gas mass retrievals using HD 112 $\mu$m and far-IR SED is less than 50\% of the true solution when SNR of the SEDs are above 15. Particularly, the low-mass disk reaches the minimum uncertainty, which is 25\% of the true solution in this study, when SNR $\geq$ 10, while the other examples show similar uncertainty at SNR $>$ 15. We have explored the uncertainty of the retrievals for similar low-mass disks and find that the uncertainty tends to be low even when the SNR is less than 10 for low-mass disks ($<$0.005 M$_\odot$). This is likely due to the lower optical depth of the disk dust structure in the far-IR SED, reducing the degeneracy of SED per given disk structures. 
}

\subsubsection{Constraining disk gas mass using velocity-resolved HD lines}

We have tested whether or not the HD emission alone may accurately constrain the disk gas mass without other information when the line profiles are spectrally resolved. We use the ProDiMo code to simulate 61 gas emission lines, including the HD emission at 56 $\mu$m and 112 $\mu$m. The velocity resolution of the HD lines is assumed to be 0.1 km/s. The velocity range of the line profiles is from -12.5 km/s to 12.5 km/s. The total number of the velocity channels is 250. While the FWHM of HD lines varies with the central star mass, inclination, and the details of HD emission, we found that a typical FWHM of the HD lines lies from 4 km s$^{-1}$ to 15 km s$^{-1}$ at the inclinations of 45 degrees (often $<$ 2 km s$^{-1}$ for 0-degree inclination), suggesting that the lines span at least 40 velocity channels. This is larger than the number of degrees of freedom in the models (number of free parameters, 15, in our study). In solving equations, one can uniquely determine the solution if there are more orthogonal/independent data points than the degree of freedom. Thus, using the HD line profiles may constrain the unique model out of the 3000 models and deliver an accurate estimate of the disk gas mass if the line profiles are relatively orthogonal to each other, and if the resolving power is sufficiently high. A highly resolved line in this context means that the number of resolution elements across the line is larger than the number of free parameters in the models. Disks have more than 15 free parameters; therefore, a fully resolved line should have $\gtrsim 15$ resolution elements across the line, which is equivalent to $R \gtrsim 1.5\times 10^6$ for a line with FWHM of 4 km s$^{-1}$.

\begin{figure*}[tb]
\centering
\includegraphics[angle=0,scale=0.7]{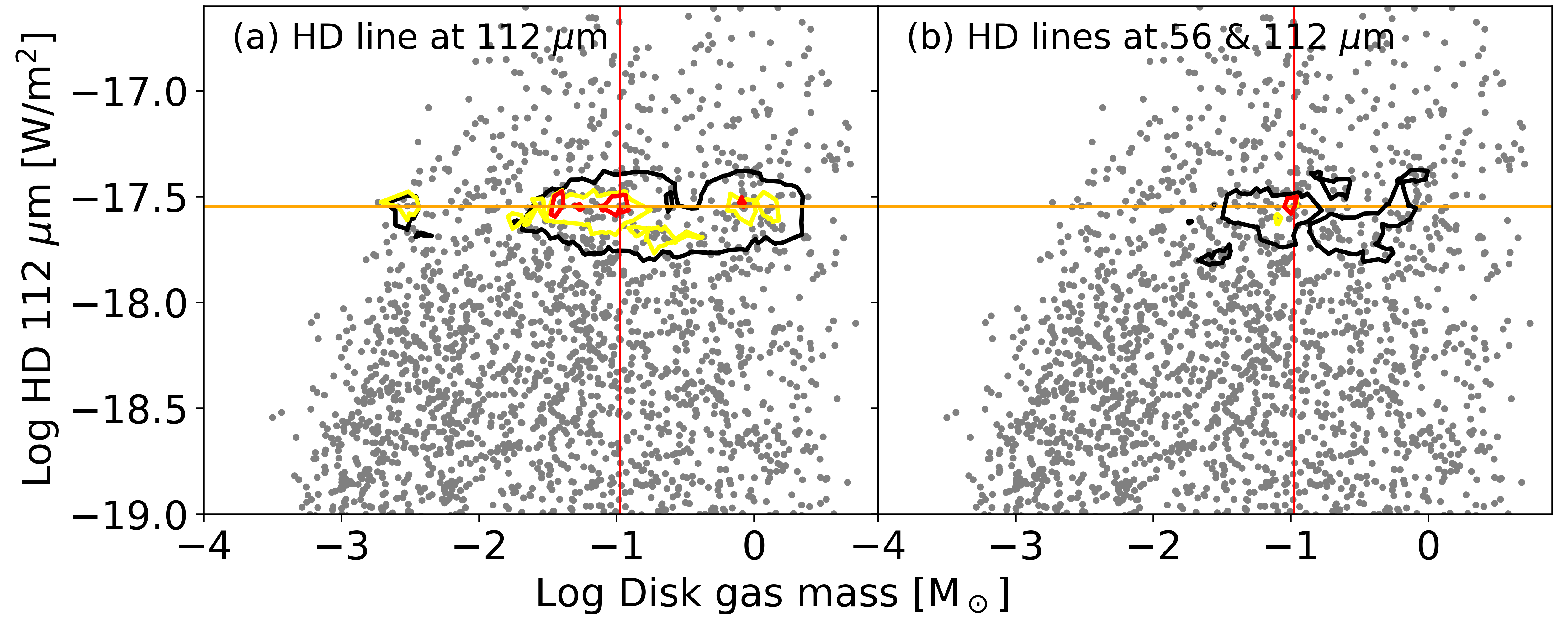}
\caption{HD flux as a function of the disk gas mass (gray dots) with the 1-$\sigma$ uncertainty contours of the HD line profile fitting. The left panel shows the uncertainties using only the HD 112 $\mu$m line, while the right panel shows the results using both HD 56 $\mu$m and 112 $\mu$m lines. The black contours are when the SNR of the peak intensity of the line is 3. The yellow and red contours denote SNR of 5 and 10, respectively. The figures show that both HD 56 and 112 $\mu$m lines are required to have a sufficiently small uncertainty in measuring the disk gas mass using line profiles. }
\label{f13}
\end{figure*}

Figure \ref{f13} shows the results of constraining the disk gas mass using the HD line profiles. We adopt Model 414 as the observed disk and calculate $\chi^2$ values of the line profile fitting using spectra from the other models. The estimation of the PDFs from the $\chi^2$ values is the same as shown in section \S\ref{sec:stats_3}. The black, yellow, and red contours denote the 1-$\sigma$ uncertainties of the PDF of fitting the HD line profiles with the SNRs of 3, 5, and 10, respectively, { at} the peak intensity { channel}. In the left panel, only the 112 $\mu$m HD line profile is used to estimate the PDFs, while both the 56 and 112 $\mu$m HD line profiles are used in the right panel. The orange horizontal line is the observed flux, and the red vertical line is the true solution of the disk gas mass. The uncertainties of the disk gas mass from the PDFs of the HD line fitting and the HD flux at 112 $\mu$m are once again determined by the minimum and the maximum intersections between the contours and the orange line. The left panel demonstrates that using only the 112 $\mu$m HD line, even with a high SNR of 10, delivers a large uncertainty of the disk gas mass, spanning roughly a factor of 50. This suggests that a single, { spectrally-resolved} HD line cannot resolve the degeneracy of models sufficiently to constrain the disk gas mass accurately. On the other hand, the right panel shows significantly smaller uncertainty, even when the lines have an SNR of 5 (intersections of the yellow contour and the orange line). Thus, the disk gas mass may be accurately determined when both HD lines at 56 $\mu$m and 112 $\mu$m are fully spectrally resolved with a SNR above 5. 

\begin{figure*}[tb]
\centering
\includegraphics[angle=0,scale=0.7]{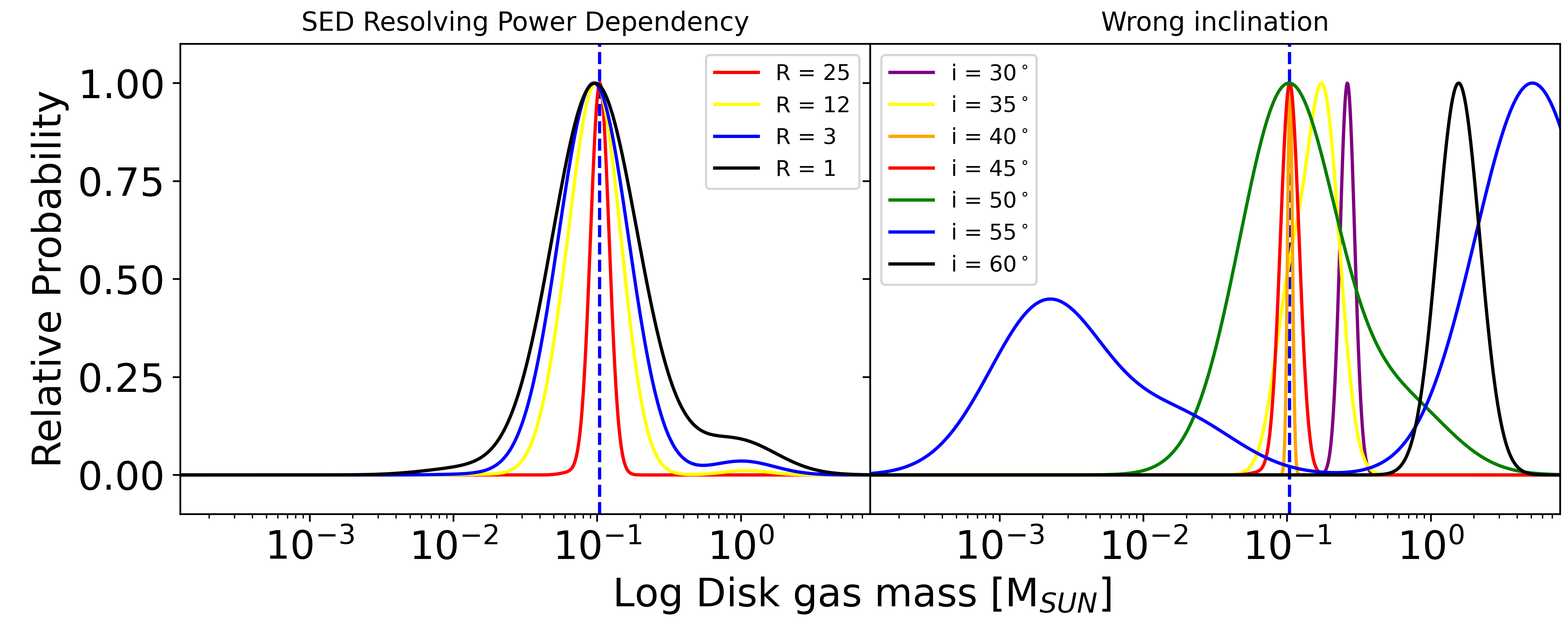}
\caption{\sout{The p}Panel (a) shows the PDFs of disk gas mass as a function of the SED resolving power. The uncertainty of the disk gas mass retrieval (the 1-$\sigma$ width of PDFs) increases with a decrease in the SED resolving power. The highest peak of PDF indicates the most probable disk gas mass. The PDFs are estimated using the same disk example in Figure \ref{f10}, Model 414, and the far-IR SED is used. Different colors of the PDFs denote different resolving power. The SNR is assumed to be 20 at each channel. The blue-dashed vertical line is the true solution. Panel (b) shows the retrieval error if the wrong inclination is assumed. Here, we show the PDFs of the disk gas mass retrievals for an assumed disk inclination of 45$^\circ$, with each colored line representing a different true inclination.  If the inclination is known to within $\pm$ 5$^\circ$ we can obtain a retrieval close to the true solution (blue-dashed vertical line), while inclination errors $>$ 10$^\circ$ dramatically increases the retrieval errors.  This suggests that the disk inclination is critical information for determining the disk gas mass using the HD 112 $\mu$m flux and the SED.  }
\label{f14}
\end{figure*}
\begin{figure*}[tb]
\centering
\includegraphics[angle=0,scale=0.7]{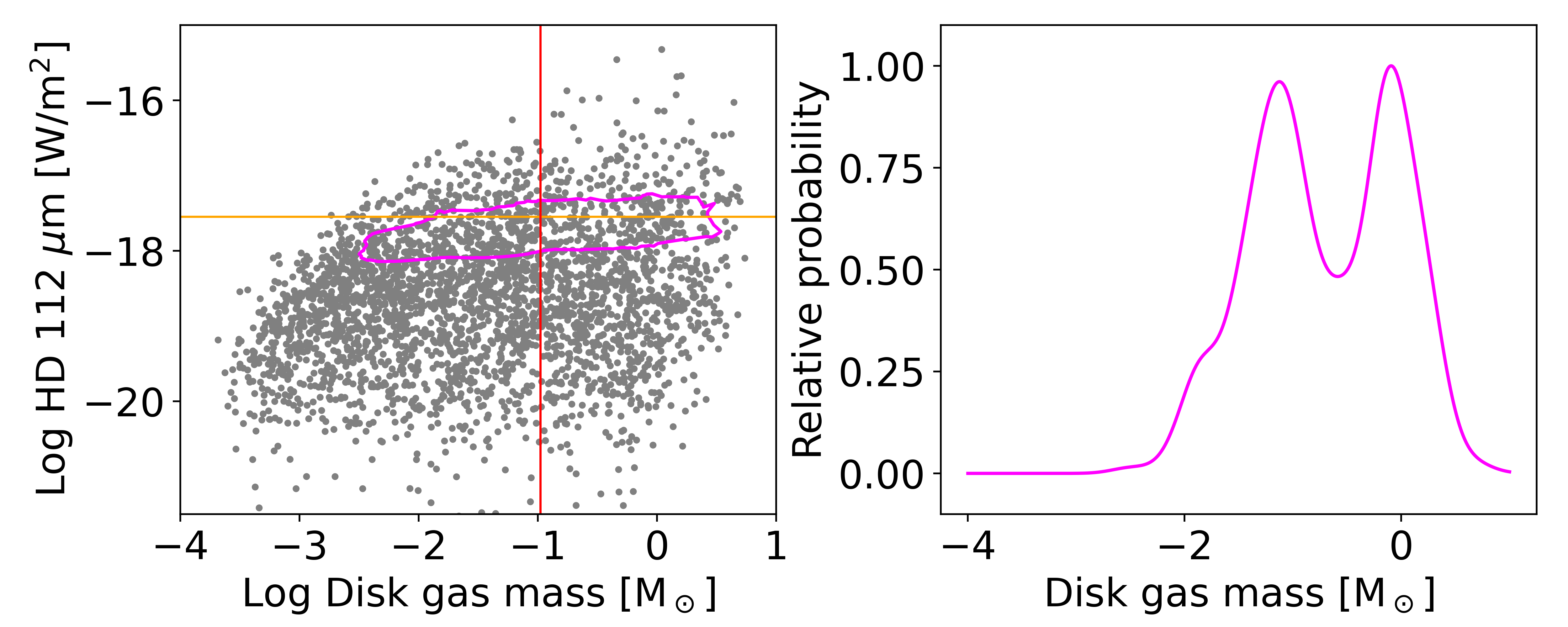}
\caption{HD flux as a function of the disk gas mass with the 1-$\sigma$ uncertainty contours of the HD flux at 56 $\mu$m. Adding the HD 56 $\mu$m flux is not adequately orthogonal to the HD 112 $\mu$m to determine the disk gas mass accurately.}
\label{f15}
\end{figure*}

\section{Discussion} \label{sec:dis}

\subsection{Limitations}

The 3000 disk models considered in this study cover significantly wider ranges of physical parameters than previous studies. However, there are still a few parameters that { have not been} explored here. One such parameter is the disk inclination. In this study, we only synthesized observables for 0 and 45 degree inclinations. Comparing the results between 0 and 45 degrees, we found that the HD flux and the SED vary significantly for disks with higher optical depths, which are typically more massive. We found that the HD flux may vary from a factor of 2 to an order of magnitude, depending on the disk mass, with a higher mass disk having larger variations. Also, the SED is very sensitive to the inclination. For example, a highly-flared disk can obscure the central star when the inclination is 45 degrees, while we may see the central star at 0 degrees. This results in vastly different SEDs in the wavelength range from UV to mid-IR. The SED from far-IR to mm wavelengths is dominated by the thermal dust emission; it is thus less affected by the inclination. The uncertainties discussed in section \S\ref{sec:stats} are estimated with fixed inclinations. The inclination should be known prior to { using} the 3000 models { to obtain} the disk gas mass with the smallest uncertainty. If the disk inclination is not known, { the error in} estimating the disk gas mass using such a statistical approach will { increase dramatically with the inclination error, as an example shown in Figure \ref{f14}. The example suggests that the inclination error should not be larger than 5$^\circ$, and the uncertainty of retrieval also increases significantly when the inclination of a disk is mistaken to be lower than the true inclination.} On the other hand, the inclination of disks may be obtained from many ALMA dust continuum images of disks, which are readily available{; thus, the inclination error may not be the critical obstacle for the statistical retrieval of the disk gas mass.}  

{ Figure \ref{f14} (left) shows the limitation of the disk gas mass retrieval when the resolving power of far-IR SED is limited. The estimation of PDFs is shown for Model 414 as a fiduciary case, the same as the panel (c) in Figure \ref{f10}, while the resolving powers of the far-IR SED are varied, as noted in the figure. At the resolving power of R = 25, the retrieval uncertainty is 25\% to the true solution, which is the minimum resolution in this study. On the other hand, the uncertainty increases to 50\% to the true solution at R = 12, a factor of two at R = 3, and a factor of three at R = 1, which suggests that at least R $>$ 12 is required to minimize the retrieval uncertainty originating from the thermochemical models. Future ALMA SEDs with a higher resolving power and far-IR space telescopes (e.g., far-IR Astrophysics Probe Explorer), which are proposed to have R $>$ 1000, will provide the required SEDs for the disk gas mass retrievals.}

We have not considered disks with gaps in this study. Dust continuum observations of disks with high spatial resolution \citep[e.g.,][]{andrews18} revealed that disks may develop many gaps in the dust distribution, while it is still unclear if the gas structure always follows the dust gaps \citep{dong17,li21,lee22}. Although ProDiMo can model disks with gaps we neglect them here since their inclusion would considerably increase the dimension of parameters and would require substantially more models to adequately explore the parameter space. If the dust gaps are not big enough to make gaps in the gas structure, the HD abundance and gas temperature may vary only slightly compared to the disk without gaps, leading to little difference in the predicted HD flux. On the other hand, since the SED reflects the dust density and temperature distribution directly, the SED will vary significantly with the inclusion of gaps \citep[e.g.,][]{liu22}. This may affect the most probable values and their uncertainties in the disk gas mass estimation using the HD flux and the SED (section \S\ref{sec:stats_4}). In such cases, having HD line profiles as additional constraints may increase the accuracy of measuring the disk gas mass and reduce the uncertainties since they only trace the gas component of the disk.

We have not explored variations in the C/O ratio and dust composition in the models. The measured C/O ratios in stars have been shown to vary within a factor of three \citep{nissen18,suarez18}. Dust composition may also vary as dust grains evolve in dense molecular environments through dust growth. Here, those parameters are fixed, and typical values based on observations of molecular clouds have been adopted. The most impacted part in the models with different C/O ratios and dust composition may again be the SED \citep{compiegne11}. Variations in the C/O ratio result in different water ice abundances on the dust grains, varying the dust opacity slightly. However, the variations in the SED with the dust composition have the strongest impact in visible and near- and mid-IR wavelengths since the solid bands are mostly in those wavelength ranges. The variation at far-IR to mm wavelengths due to different dust compositions is rather small.

\subsection{HD emission at 56 $\mu$m}

We mainly explored the statistics of the HD emission at 112 $\mu$m along with other observables in section \S\ref{sec:stats}. But the 3000 models also include the 56 $\mu$m HD flux (Figure \ref{f3}) and many other observables (61 emission lines in total). \citet{trapman17} reported that the 56 $\mu$m HD flux may be complementary to the 112 $\mu$m HD flux in estimating the disk gas mass within their models. We also probed whether or not the 56 $\mu$m HD flux may work as an orthogonal constraint to the HD flux at 112 $\mu$m when the information about the target disk is minimal. Figure \ref{f15} shows the 1-$\sigma$ uncertainty contour of the 56 $\mu$m HD flux. We assumed that the 56 $\mu$m HD flux is observed to be 1.99$^{+1.9}_{-0.99}$ $\times$ 10$^{-20}$ W m$^{-2}$. The contour is elongated parallel to the orange line, which is the observed HD flux at 112 $\mu$m. This demonstrates that the 56 $\mu$m HD flux is not sufficiently orthogonal to the 112 $\mu$m HD flux to significantly reduce the uncertainty in constraining the disk gas mass.

\section{Summary \& Conclusions} \label{sec:sum}

The gas content of protoplanetary disks constrains various processes { involved in} planet formation. Unfortunately, probing the gas component has been very challenging due to the physical limitations of directly observing H$_2$ molecules, which are the dominant component of the gas. HD, the singly-deuterated counterpart of H$_2$, has been suggested as the best option to trace the distribution of H$_2$. Many studies have reported that the HD flux may be a promising tracer for estimating the disk gas mass. However, the physical parameters that have been explored are limited, leaving doubts as to whether HD is a good tracer in real-life { situations}, where we may not fully constrain all disk parameters. In this study, we expanded the previous studies by carrying out extensive disk modeling with 15 free parameters. We modeled  3000 disks with very wide parameter ranges and studied the relationship between observables, including the HD flux and the dust SED, and the physical parameters, { including the} disk mass. We summarize our findings as follows:   
\begin{enumerate}

\item The HD fluxes from the full set of 3000 models do not show a strong correlation with the disk gas mass, while there may be an upper limit of the HD flux correlating with the disk gas mass. This suggests that the disk gas mass cannot be estimated reliably if the HD flux is the only information about the target { (which is rarely the case)}.

\item HD fluxes at 112 microns for disks at 140 pc typically are in the range of 10$^{-19}$ to 10$^{-17}$ W m$^{-2}$. To conduct a complete HD survey of disks at $\leq$140 pc (Taurus and Ophiuchus molecular clouds), a telescope should have a sensitivity of { at least} $\sim$10$^{-19}$ W m$^{-2}$.

\item The { mass-averaged} HD abundance is almost constant, with variations less than a factor of two, even when the total UV flux from the central star and the ISRF vary by five orders of magnitude. This is because the UV shielding by H$_2$ is highly efficient in disks and confirms the results of previous studies \citep[e.g.,][]{trapman17,kama20}.  

\item  The HD flux is sensitive to the radiative transfer processes, including excitation conditions and optical depth. We find that the 112 $\mu$m HD flux is closely related to the mass-averaged gas temperature for low-mass/optically thin disks. On the other hand, the HD flux of massive disks shows { a} large scatter for a given gas temperature.   

\item Probing the key physical information required to constrain the disk gas mass using the HD flux, we find that knowledge of at least four physical parameters referring to the central star, { the} disk size, { the} dust properties, and { the} disk inclination { are} required. With the four parameters known, we may determine the disk gas mass within a factor of three.  

\item  Using the far-IR or mm/sub-mm SED along with the HD flux allows to constrain the disk gas mass with sufficiently small uncertainty (less than a factor of two) when the SNR of the SED is higher than 15. The inclination affects the SED and HD flux significantly, particularly for 
massive disks (M$_{gas}$ $>$ 0.1 M$_\odot$). Thus, inclination should be accurately measured to determine the disk gas mass using the HD flux and the SED. The distance should also be known.

\item  Using fully spectrally-resolved 56 $\mu$m and 112 $\mu$m HD line profiles may constrain the disk gas mass with small uncertainty (less than a factor of two), provided that the resolving power is $R \gtrsim 1.5\times 10^6$. 
\end{enumerate}

The key result of this study is that we may estimate the disk gas mass with adequate uncertainty directly from HD flux and { far-IR/sub-mm} SED, or from HD line profiles alone without knowing too many details about the disks. Our models have their own limitations since there are certainly more parameters that are not explored (e.g., gaps, C/O ratio, etc.). However, within its limitations, the statistical approach in this study shows a promising way to make the first-order estimate of the disk gas mass from the far-IR observations.

\software{ProDiMo \citep{woitke09,kamp10}, scitkit-learn \citep{pedregosa11}, Miex \citep{wolf18} }

\begin{acknowledgments}
This research was carried out at the Jet Propulsion Laboratory, California Institute of Technology, operated under a contract with the National Aeronautics and Space Administration (80NM0018D0004). Funding was provided in part by a Strategic Initiative from the Engineering and Science Division of JPL.
\end{acknowledgments}

%

\vspace{5mm}

\bibliography{disk_HD_ref}{}
\bibliographystyle{aasjournal}



\appendix

\section{Correlation of physical parameters to the HD 1$-$0 flux}

\begin{figure*}[tb]
\centering
\includegraphics[angle=0,scale=0.85]{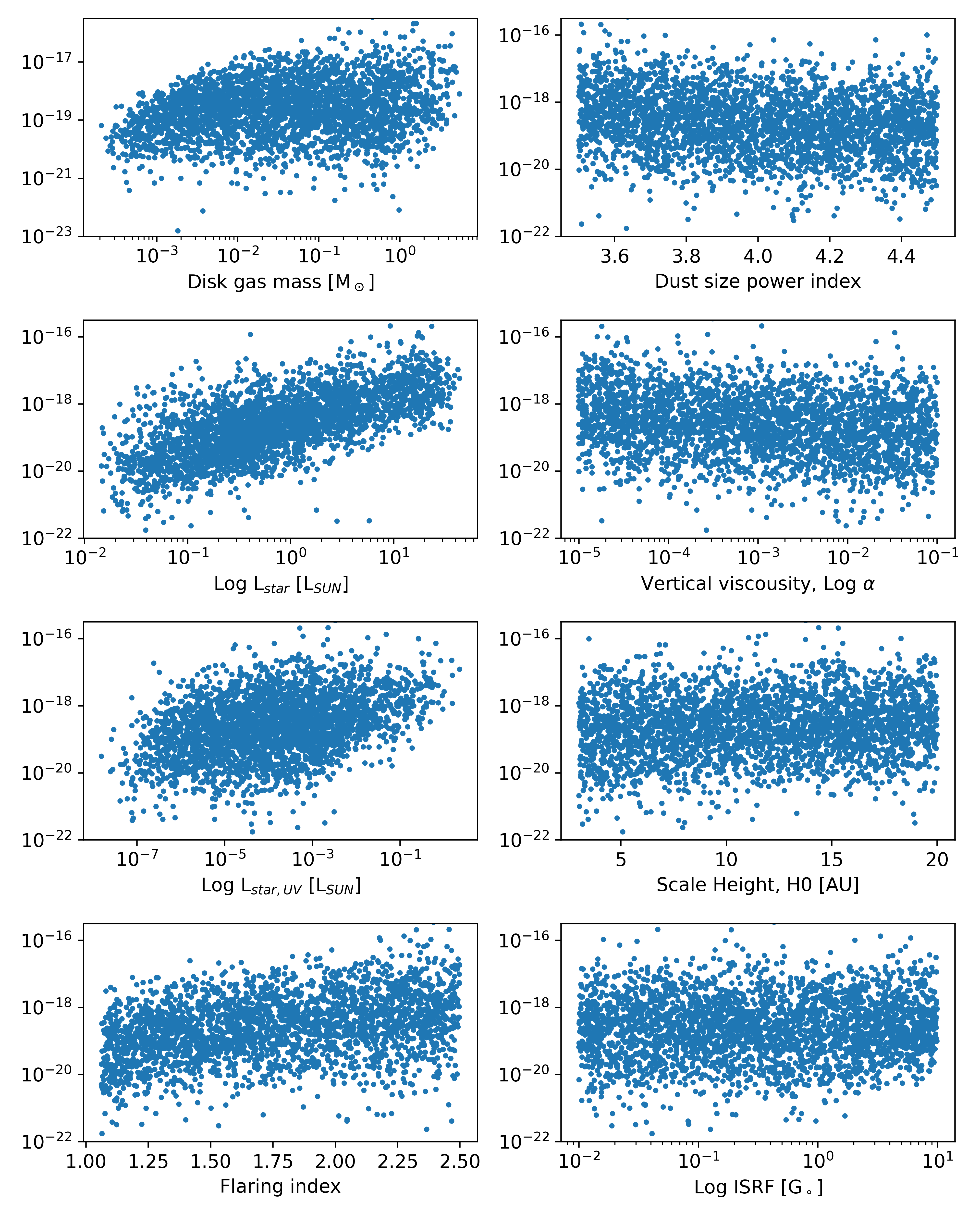}
\caption{ HD flux as a function of eight different input parameters. The vertical axis is the HD 112 $\mu$m flux in W m$^{-2}$ on a logarithmic scale.}
\label{fA1}
\end{figure*}

{ Figure \ref{fA1} shows the correlation of eight input parameters to HD 112 $\mu$m flux. As in the previous studies \citep[e.g.,][]{trapman17,kama20}, the HD 112 $\mu$m flux is sensitive to the gas temperature. A hotter and more massive disk tends to emit { greater} HD 112 $\mu$m flux; thus, the parameters that determine gas temperature show a relatively stronger correlation to HD 112 $\mu$m flux. In the figure, the luminosity of the central star shows the strongest correlation { with} HD 112 $\mu$m flux since it is the main source of heating the disk gas. The stellar UV luminosity also has a relatively strong correlation to HD 112 $\mu$m flux. The disk gas mass clearly shows a correlation to the upper limit of HD 112 $\mu$m flux; however, there is a large scatter toward lower HD fluxes since there are diverse variations of temperature structure even for the same disk mass. The parameters that define disk gas distribution also show correlations to HD 112 $\mu$m flux. The trend shows that more flared disks tend to have brighter HD 112 $\mu$m flux, which is possibly due to absorbing more photons from the central star and having less optical depth of HD 112 $\mu$m transition. { The} HD 112 $\mu$m flux is also affected by the dust size distribution. With a higher dust size distribution power index (more small dust { grains} for a given dust mass), the HD 112 $\mu$m flux tends to decrease. This is because the larger number of smaller dust increases the continuum optical depth of the disks, making the disks colder. A similar effect is also seen with the vertical viscosity. A larger vertical viscosity indicates that the dust settling is slower, and more dust grains remain { in} the upper part of the disks. This makes the colder region of a disk larger, which essentially reduces the collisional excitation { rate} of HD.   

There are seven other input parameters for the models. However, we do not show them in the figure since their correlations to the HD 112 $\mu$m flux are relatively minor than the eight parameters shown in the figure.}

\section{Disk Physical Structure}

{
We use parametrized disk structure in this study to explore a wide range of disk structures that cannot be simulated under hydrostatic assumption due to unknown physics. The shape and mass distribution of the gas in the disks follows the below equation, which is also shown in \citet{woitke16}.
\begin{equation}
    \Sigma(r)~\propto~r^{-\gamma}\exp{\left(-{r\over R_{Taper}}\right)} ,
\end{equation}
where $\Sigma$ is the surface density, $r$ is radius, $R_{Taper}$ is tapering off radius. $R_{Taper}$ is fixed to be 0.75$\times$R$_{out}$. The vertical gas structure follows a Gaussian gas distribution as below:
\begin{equation}
    \rho(r,z) \propto \exp{-{z^2\over 2{\rm H}(r)^2}},
\end{equation}
where $\rho$ is gas density, $r$ is the radius, $z$ is vertical height, and H$(r)$ is the scale height. The H$(r)$ is defined as  
\begin{equation}
    H(r)~=~{\rm H_0} \left(r \over r_0\right)^\phi,
\end{equation}
where $r_0$ is the reference radius, and H$_0$ is the scale height at the reference radius. In this study, the reference radius is 100 AU. 

The dust size distribution is assumed with the power law index,
\begin{equation}
    dn(a)~\propto~a^{-\beta}da,
\end{equation}
where $n$ is number density of dust grains, $a$ is the dust radius, and $\beta$ is the dust size distribution power index.
}

\end{document}